\documentclass[useAMS,usenatbib]{mn2e}
\usepackage{ifpdf}
\ifpdf
\end{document}
\fi

\usepackage{amsmath,fleqn,graphicx,amssymb,xcolor}
\usepackage{multirow}
\defcitealias{JJBPJM16}{BM16}
\renewcommand{\[}{\begin{equation}}
\renewcommand{\]}{\end{equation}}
\def\p{\partial}\def\i{{\rm i}}

\def\Rsh{R_{\rm sh}}
\def\Jcrit{J_{z\rm crit}}
  \def\Js{J_{\rm s}}

\def\agamaTwo{{\sc agama}b}
\def\agama{{\sc agama}}
\def\los{{\sc los}}

\def\rd{}

\let\boldgrk=\gkvecten
\let\boldgrksc=\gkvecseven

\def\gkthing#1{{\mathchoice%
	{\hbox{{\boldgrk\char#1}}}
	{\hbox{{\boldgrk\char#1}}}
	{\hbox{{\boldgrksc\char#1}}}
	{\hbox{{\boldgrksc\char#1}}}}}

\def\mapleftright#1{\smash{\mathop{\longleftrightarrow}\limits^{#1}}}

\def\vtheta{\gkthing{18}}

{\newif\ifnotend
\notendtrue
\def\veclist{ABCDEFGHIJKLMNOPQRSTUVWXYZabcdefghijklmnopqrstuvwxyz.}
\def\top#1#2.{#1}
\def\tail#1#2.{#2.}
\loop\expandafter\xdef\csname v\expandafter\top\veclist\endcsname%
{{\noexpand\bf\expandafter\top\veclist}}
\edef\veclist{\expandafter\tail\veclist}
\if\veclist.\notendfalse\fi\ifnotend\repeat}
{\newif\ifnotend
\notendtrue
\def\veclist{ABCDEFGHIJKLMNOPQRSTUVWXYZ.}
\def\top#1#2.{#1}
\def\tail#1#2.{#2.}
\loop\expandafter\xdef\csname c\expandafter\top\veclist\endcsname%
{{\noexpand\cal\expandafter\top\veclist}}
\edef\veclist{\expandafter\tail\veclist}
\if\veclist.\notendfalse\fi\ifnotend\repeat}
\def\d{{\rm d}}

\def\bolOm{\mbox{\boldmath$\Omega$}}
\def\vOmega{\bolOm}

\def\Gyr{\,\mathrm{Gyr}}
\def\Myr{\,\mathrm{Myr}}
\def\kpc{\,\mathrm{kpc}}

\def\kms{\,\mathrm{km\,s}^{-1}}

\def\masyr{\,\mathrm{mas\,yr}^{-1}}

\def\e{\mathrm{e}}

\def\fracj#1#2{{\textstyle{#1\over#2}}}

\def\rms{\textsc{rms}}
\def\vthetaT{{\vtheta^{\rm T}}}\def\vJT{{\vJ^{\rm T}}}
\def\thetaT{\theta^{\rm T}}\def\JT{J^{\rm T}}

\title[A New Torus Generator for AGAMA]
{A New Torus Generator for AGAMA}

\author[James Binney, Thomas J Wright \& Eugene Vasiliev]{
  James Binney$^1$\thanks{E-mail: binney@physics.ox.ac.uk}   Tom Wright$^{2}$
  and Eugene
  Vasiliev$^{3}$\\  
  $^1$Rudolf Peierls Centre for Theoretical Physics, Clarendon Laboratory,
  Oxford, OX1 3PU, UK\\
  $^2$Somerville College, Oxford OX1\\
  $^3$Department of Physics \& Astronomy, University of Surrey, Guilford GU2
  7XH
}

\begin{document}
\maketitle

\begin{abstract}
Code is presented that computes and exploits orbital tori for any
axisymmetric gravitational potential. The code is a development of the
{\sc agama} software package for action-based galaxy modelling and can be
downloaded as the \agamaTwo\ code library. Although coded in
C++, most of its functions can be accessed from Python.  We add to the
package functions that facilitate confronting models with data, which involve
sky coordinates, lines of sight, distances, extinction, etc. The new torus
generator can produce tori
for both highly eccentric and nearly circular orbits that lie beyond the
range of the earlier torus-mapping code.  Tori can be created by
interpolation between tori at very low cost. Tori are fundamentally
devices for computing ordinary phase-space coordinates from angle-action
coordinates, but \agamaTwo\ includes an action finder that returns angle-action
coordinates from any given phase-space location. This action finder yields
the torus through the given point, so it includes the functionality of an
orbit integrator. The action finder is more accurate and reliable but
computationally more costly than the widely used St\"ackel Fudge. We show how
\agamaTwo\ can be used to generate sophisticated but cheap models of tidal streams
and use it to analyse data for the GD1 stream. With the most recently
published distances to the stream, energy and angular momentum imply that the
end that must be leading is trailing, but extremely small changes to the
distances rectify the problem.
\end{abstract}


\section{Introduction} \label{sec:intro}

AGAMA (Action-Based Galaxy Modelling Architecture, \citealt{agama}) is a
suite of programs with which to model galaxies like ours. With {\sc agama}
one can quickly recover the gravitational potential that is self-consistently
generated by a specified density of gas in conjunction with dark matter and
several populations of stars.  Stars and dark-matter particles are specified
by distribution functions $f(\vJ)$ that depend on the action integrals $J_r$,
$J_z$ and $J_\phi$. The suite can then be used to examine the model's
dynamics and simulate observations of the model galaxy, either by an observer
internal to the model, or by an external observer. The suite also includes
code for fitting a mean-field potential $\Phi(R,z)$ to an N-body model. 

Action-based modelling depends on algorithms for shifting between
angle-action coordinates $(\vtheta,\vJ)$ and conventional phase-space
coordinates $(\vx,\vv)$.  AGAMA uses the St\"ackel Fudge
\citep{JJB12:Stackel} for transformation $(\vx,\vv)\to(\vtheta,\vJ)$ and
torus mapping \citep{McGJJB90,KaJJB94:MNRAS} for the inverse transformation
$(\vtheta,\vJ)\to(\vx,\vv)$. The code \cite{agama} included in \agama\ for
torus mapping is a superficial wrapping of the code released by Binney \&
McMillan (2016; hereafter \citepalias{JJBPJM16}).\footnote{The agama\ code
differed from the BM16 code by using the algorithm for angle finding that is
described in Section~\ref{sec:solving} below.} Much of this code dates back
to work by W.~Dehnen in 1995/6, and was itself based on the 1994 doctoral
thesis of M.~Kaasalainen. The code was later augmented to facilitate
perturbation theory \citep{Binney2016,Binney2018,Binney_negJ}. Given the
code's age, it is natural that it is not as clean or as fast as it could be.
Moreover, its classes do not coincide with classes that are basic in {\sc
agama}, which generates unnecessary complexity and confusion. For these
reasons we have undertaken a complete re-write of the torus-mapping code to
produce a code that is native to {\sc agama} and works faster and more
reliably. In the course of this endeavour we reconsidered aspects of the way
the old code worked and made major changes to the algorithms. These changes
enable the new code to produce tori for highly eccentric orbits that lay
beyond the scope of the old code.  

{\rd
Tori make it possible to investigate resonant phenomena via Hamiltonian
perturbation theory \citep{Binney2016}.  Crucial to this application is the
ability to produce tori by interpolating on a grid of tori. The Torus Mapper
released by BM16 was implemented in such a way that in some circumstances
interpolation fails. The changes reported here facilitate interpolation
between tori and thus applications of perturbation theory. We have also taken
the opportunity to supplement the dynamical modelling code with code that
facilitates comparisons between models and observational data. 

The extensions reported here are embedded in a fork from  the code released
by \cite{agama} that would not be easy to merge with the current release of
\agama. Hence we are releasing as \agamaTwo\ an independent branch of \agama.
Fortunately, most of the changes to \agama\ functions are transparent to
users.

While tori produce a map $(\vtheta,\vJ)\to(\vx,\vp)$, the inverse map is the
more useful for stellar dynamics. This is generally provided by the St\"ackel Fudge
\citep{JJB12:Stackel}, which has a weak conceptual foundation and sometimes
performs
badly \citep{WrightB}. AGAMAb contains a more reliable and capable, but
much more costly, action finder for the inverse mapping
$(\vx,\vp)\to(\vtheta,\vJ)$.
}

Orbits are the core concepts of both stellar and planetary dynamics, and they
are central to Schwarzschild modelling, which is used extensively in the
interpretation of observations of external galaxies
\citep[e.g.][]{ATLAS3D_short,Cappellari2016,Zhu2018,CALIFA2024,MAGPI2024}.
A torus is in many ways an extension of the idea of an orbit and it seems
probable that replacing orbits by tori will make Schwarzschild modelling an
even more powerful technique.

Significant fractions of dark-matter particles and the stars of our Galaxy's
stellar halo are on highly eccentric tori that the BM16 code cannot
construct. \cite{WrightB} show that highly eccentric orbits should be divided
into two classes according as their action $J_z$ is less than or greater than
a critical action $\Jcrit(E)$. The new torus generator is able to create tori
for almost all orbits by using different `toy maps' when $J_z$ lies either
side of $\Jcrit$ when $|J_\phi|$ is small.

{\rd
Section \ref{sec:fundamentals} lays out the conceptual framework and the
basic equations. Section \ref{sec:inBox} explains how to use the code by
plotting diagrams that illustrate its power. Section \ref{sec:streams} uses
the code to model tidal streams. Section \ref{sec:Python} describes the
Python interface. Section \ref{sec:conclude} summarises and discusses
possible applications.  Appendix \ref{sec:behindScenes} explains in some
detail how the code creates tori using new algorithms for choosing HJ maps
and defining point transformations. It also explains interpolation between
tori and the use of diagnostic plots that showcase the utility of point
transformations.  Appendix \ref{sec:obs} describes functions in \agamaTwo\
that relate dynamical quantities to observable ones, such as sky coordinates,
line-of-sight velocities and magnitudes.
}

\section{Fundamentals}\label{sec:fundamentals}

A torus $T$ is an injective map $T(\vtheta)\to(\vx,\vv)$ from
three-dimensional angle space into six-dimensional phase space. $T$ is $2\pi$
periodic in each component of the three-dimensional vector $\vtheta$. Tori
are labelled by their three actions
\[
J_i={1\over2\pi}\int_0^{2\pi}\d\theta_i\,\vp\cdot{\p\vx\over\p\theta_i}
\] 
where $(\vx,\vp)$ are the image points associated with $\vtheta$. The values of the
integrals depend neither on the values to which we set the angles not
integrated over, nor on which canonical coordinate system we use for
$(\vx,\vp)$.

If all image points $(\vx,\vv)$ lie on a level surface $H(\vx,\vv)=E$ of the
Hamiltonian $H$, the torus is equivalent to an orbit in the sense that given
a vector of angles $\vtheta_0$ and a vector of frequencies $\vOmega$, the
time series
\[
[\vx(t),\vv(t)]=T(\vtheta_0+\vOmega t),
\]
 with $\Omega_i=\p H/\p J_i$ solves Hamilton's equations of motion for
$H$. 

We call a torus that lies within a constant-Hamiltonian surface an {\it
orbital torus}. With such a torus we associate values of the vector $\vOmega$
and the scalar $E$ in addition to vector $\vJ$.

The classical procedure for obtaining  tori is solution of the Hamilton
Jacobi equation
\[
H\Big(\vx,{\p S\over\p\vx}\Big)=E
\]
 for the generating function $S(\vx,\vJT)$ of the canonical transformation
$(\vx,\vp)\leftrightarrow(\vthetaT,\vJT)$, where the superscripts T stand
for `toy' to distinguish this angle-action system from the one,
$(\vtheta,\vJ)$, required by orbits in the given, real potential.

The Hamilton-Jacobi equation can be solved for very
few Hamiltonians, and we start from one of these: for orbits that
are not highly eccentric, we use the isochrone potential
\citep{Henon1959,Henon1960}
\[\label{eq:PhiIso}
\Phi(r)=-{\Js^2/b\over{b+\sqrt{b^2+r^2}}},
\] 
 while for highly eccentric orbits (ones with $|J_\phi|/J_r$ small and
$J_z/J_r\la1$), we use the
effective potential of a semi-degenerate three-dimensional harmonic
oscillator
\[\label{eq:PhiHO}
\Phi(R,z)=\fracj12\big(\omega_R^2 R^2+\omega_z^2z^2\big).
\] 
 In equation (\ref{eq:PhiIso}), $b$ is a length scale and $\Js=\sqrt{GMb}$ is a
scale action, with $M$ the mass of the body that generates the potential, and
$G$ is Newton's constant. In equation (\ref{eq:PhiHO}), $\omega_R$ and $\omega_z$
are suitable constants and $R=\sqrt{x^2+y^2}$ is cylindrical radius. For
specified $\vJT$ and either choice of potential, analytic formula give
$(\vx,\vp)$ as functions of $\vthetaT$. Other formulae yield $(\vthetaT,\vJT)$
given $(\vx,\vp)$ \citep[see][\S3.5.2]{GDII}. We call a map provided by an
isochrone or a harmonic oscillator an `HJ map'. 

Since an isochrone potential is spherical, all its orbits conserve total
angular momentum $L\equiv J_z+|J_\phi|$, and they are excluded from a region
around the
origin by a contribution $\frac12L^2/r^2$ to the effective potential. In a
non-spherical potential, $J_z$ contributes to the analogous centrifugal
barrier only above a critical value $\Jcrit$ \citep{WrightB}. When
$J_z<\Jcrit$, orbits are better modelled by harmonic-oscillator orbits
than by  isochrone orbits. One way to understand this is to recall that in a
semi-degenerate harmonic oscillator, oscillations in $z$ are independent of
oscillations in $(R,\phi)$ while in a spherical potential such as the
isochrone oscillations in $z$
largely reflect the radial and azimuthal oscillations in the orbit's inclined
plane.

\subsection{Point transformations}

We combine an HJ map with a point transformation
$\vx\leftrightarrow\vX\equiv(R,z,\phi)$.  With $P_i$ the momentum conjugate
to $X_i$, the generating function of a point transformation is
\[\label{eq:GFpt}
\cG(\vx,\vP)=\sum_i X_i(\vx)P_i.
\]
The derivatives of $\cG$ yield the mapping between the momenta of the two
systems. The choice of point transformation depends on which HJ map has been
chosen and we defer details to Appendix~\ref{sec:PtTrans}. Here we note that
the map always involves a confocal ellipsoidal coordinate system that is
defined by the $z$ coordinates $\pm\Delta$ of its foci. 

{\rd The majority of orbits require an isochrone HJ map, and in this case the
$\Delta$ is chosen by reference to a shell orbit $J_r=0$. Shell orbits form a
two-parameter family: \agamaTwo\ takes the parameters to be either $(E,\xi)$
or $(L,\xi)$, where $E$ is energy and $L\equiv J_z+|J_\phi|$, while
$\xi=|J_\phi|/L$.  When a potential
is initialised, \agamaTwo\ computes a grid of shell orbits, and, by fitting
to
these the ellipse
 \[\label{eq:ellipse}
{R^2\over\Rsh^2}+{z^2\over\Rsh^2+\Delta^2}=1,
\]
tabulates $\Delta$ as a function of both
$(E,\xi)$ and $(L,\xi)$. At the same time, \agamaTwo\ tabulates $\Rsh$, which is the
value of $R$ at which the shell orbit crosses $z=0$.  When \agama\
initialises an {\it actionFinder} that uses the St\"ackel Fudge, it computes
an equivalent grid of shell orbits. The differences between the two versions
of the code are (i) \agamaTwo\ attaches the tables to an instance of a
{\it BasePotential} rather than an {\it actionFinder} and (ii) \agamaTwo\ stores the
variables as functions of $(L,\xi)$ in addition to $(E,\xi)$.

Mapping orbits with small $J_r$ requires that the point transformation maps
inclined circular toy orbits into shell orbits in full phase space, not
merely in real space \citep{KaJJB94:MNRAS}. Hitherto this has been
accomplished by solving o.d.e.s for two functions $\eta(r)$ and
$\zeta(\vartheta)$ as described in Appendix A3 of \cite{JJBPJM16}. This
procedure defines the mapping only in a restricted range of the meridional
plane and there isn't a natural way to extend it to the whole plane (see
Fig.~A4 in \citealt{JJBPJM16}). AGAMAb solves the problem with two
Fourier series, one each for the radial and angular outputs of the HJ map --
see Appendix \ref{sec:behindScenes} for details. Each Fourier series has five
coefficients, bringing the parameter count for a point transformation to 11.
The mapping produced this way automatically covers the entire meridional
plane and is defined by parameters that lend themselves to interpolation,
which is important because much of the power of tori for galaxy
modelling derives from the ability to obtain new tori by interpolating
between old ones.
}

The mapping
\[
(\vJT,\vthetaT)\quad\mapleftright{{\rm HJ\, eqn}}\quad(\vx,\vp)\quad
\mapleftright{\cG}\quad(R,z,\phi,p_R,p_z,p_\phi)
\]
 is canonical because it is a compound of two canonical maps. We call this
compound map the `toy map'.

To secure the flexibility to fit our image tori closely to level surfaces of
any galactic Hamiltonian
\[\label{eq:defH}
H=\fracj12\Big(p_R^2+p_z^2+{p_\phi^2\over R^2}\Big)+\Phi(R,z),
\]
we precede the toy map with the canonical transformation\[
(\vtheta,\vJ)\quad\mapleftright{S}\quad(\vthetaT,\vJT)
\]
that has generating function
\[\label{eq:defS}
S(\vJ,\vthetaT)=\vJ\cdot\vthetaT+\sum_\vk S_\vk(\vJ)\sin(\vk\cdot\vthetaT),
\]
where $(\vtheta,\vJ)$ are (approximations to) the angle-action coordinates of
the full galactic Hamiltonian. The subscripts $\vk$ in equation (\ref{eq:defS}) are
three-dimensional vectors with integer components, but in the axisymmetric
case $S(\vJ)$ is non-zero only when $k_\phi=0$. Differentiating equation (\ref{eq:defS})
we get
\[\label{eq:JT2J}
\vJT=\vJ+\sum_\vk\vk S_\vk\cos(\vk\cdot\vthetaT),
\]
 so for any chosen $\vJ$ and coefficients $S_\vk$, $\vJT$ becomes a function
of $\vthetaT$ and then with the toy map
$(\vthetaT,\vJT)\leftrightarrow(R,z,\phi,p_R,p_z,p_\phi)$, the ordinary phase
space coordinates become functions of $\thetaT$. The Marquardt-Levenberg
routine \citep[e.g.][]{NumRec} is now used to adjust the $S_\vk$ to minimise
the variance in $H(R,z,\phi,p_R,p_z,p_\phi)$ over a regular grid in
$\thetaT$.

\cite{McGJJB90} introduced the use of a generating function (\ref{eq:defS})
and the isochrone HJ map. \cite{KaJJB94:MNRAS} showed that a point
transformation was required for orbits with small $J_r$. The torus mapper
published by \cite{JJBPJM16} uses a point transformation only when $J_r/J_z$
is small, and interpolation between tori with and without point
transformations is problematic.  \cite{Ka95:closed} used
harmonic-oscillator HJ maps but to model resonantly trapped orbits rather
than untrapped but highly eccentric orbits. The reasoning that leads us to
use harmonic-oscillator HJ maps for highly eccentric orbits is given in
\cite{WrightB}.

\subsection{Solving for frequencies and angles}\label{sec:solving}

The manner in which we solve for the true angle variables $\vtheta(\vthetaT)$
is another material respect in which the new torus mapper differs from the
BM16 version. From (\ref{eq:defS}) we have
\[\label{eq:thT2th}
\vtheta=\vthetaT+\sum_\vk{\p S_\vk\over\p\vJ}\sin(\vk.\cdot\vthetaT).
\]
When $\vthetaT$ is required given $\vtheta$ it is recovered from this
equation by Newton-Raphson
iteration starting from $\vthetaT=\vtheta$.

We now derive the algorithm used to find the gradients $\p S_\vk/\p J_i$ in
detail because both the derivation given in \cite{LaaksoKaas} and the derivation
of the closely related algorithm in \cite{JJBKu93} lack clarity about the
nature of partial differentials.

$H$ is a function of only three variables, usually taken to be the components
of $\vJ$, but at any fixed value of $\vthetaT$, equation (\ref{eq:JT2J}) makes $\vJ$ a
function of $\vJT$, so we may write
\begin{align}
\Omega_i&=\sum_j\left({\p H\over\p\JT_j}\right)_\vthetaT
\left({\p\JT_j\over\p J_i}\right)_\vthetaT\cr
&=\sum_{j}\left({\p H\over\p\JT_j}\right)_\vthetaT
\left(\delta_{ij}+\sum_\vk k_j{\p S_\vk\over\p J_i}\cos(\vk\cdot\vthetaT)\right).
\end{align}
Rearranging we get
\[\label{eq:rearranged}
\Omega_i-\sum_j\left({\p H\over\p\JT_j}\right)_\vthetaT\sum_\vk k_j{\p
S_\vk\over\p J_i}\cos(\vk\cdot\vthetaT)
=\left({\p H\over\p\JT_i}\right)_\vthetaT.
\]
During the minimisation process we have already computed on a grid of points
over the torus the values of
\[
{\p H\over\p S_\vk}=\left({\p H\over\p\vJT}\right)_\vthetaT\!\!\cdot{\p\vJT\over\p S_\vk}=
\sum_j\left({\p H\over\p\JT_j}\right)_\vthetaT\! k_j\cos(\vk\cdot\vthetaT).
\] 
After using this expression to simplify the term in (\ref{eq:rearranged}) with a
cosine, we obtain
 \[\label{eq:giveOmega}
\Omega_i-\sum_\vk{\p H\over\p S_\vk}\cdot{\p S_\vk\over\p J_i}=\left({\p
H\over\p\JT_i}\right)_\vthetaT.
\]
This is a linear equation for the unknowns $\Omega_i$ and $\p S_\vk/\p J_i$
with coefficients $\p H/\p S_\vk$ and $\p H/\p\vJT$ that have already been
computed when solving for the $S_\vk$. It should hold at any point on the
torus, and in particular at all of our grid points. So if we have more grid
points than three plus the number of $S_\vk$ under consideration, we have an
over-determined system, which we solve by minimising the sum over a grid of
toy angles of the squared differences between the two sides of the equation.

{\rd
The BM16 mapper obtained the derivatives of $S_\vk$ and the frequencies by
solving a system of linear equations, but it obtained the equations'
coefficients by integrating Hamilton's equations for short time intervals
from initial conditions set by the torus on a sparse grid in $\thetaT$. Tests
showed that the two methods produce equivalent results but the new method is
$\sim30$ percent faster and makes for more compact code.

\subsection{A torus-based action finder}

Fundamentally torus mapping provides the  map from angle-action to ordinary
coordinates. The standard inverse mapping is provided by the St\"ackel Fudge
\citep{JJB12:Stackel}. Although widely used, the Fudge has a weak conceptual
underpinning and in some circumstances it can fail badly
\citep{WrightB}, so we have added to \agamaTwo\ an action finder
that adjusts the
$(\vtheta,\vJ)$ values that are returned by the St\"ackel Fudge for a given
phase-space point $\vw$.  The natural algorithm is a Newton-Raphson search of
$(\vtheta,\vJ)$ space. For such a search one needs to be able to compute
$\p\vw/\p\vtheta$ and $\p\vw/\p\vJ$.  Computing derivatives w.r.t.\ $\vtheta$
is straightforward but derivatives w.r.t.\ $\vJ$ are problematic unless the
toy map is held constant -- the algorithm used to choose the toy map makes it
hard to compute the derivatives of the map's parameters $\Delta,\Js,b$, etc.\
w.r.t.\ $\vJ$. For fixed toy map we have
\[
{\p\vw\over\p\vJ}={\p\vw\over\p\vJT}\cdot{\p\vJT\over\p\vJ}
\]
 with both the matrices on the right available. However, once the toy map has
been fixed, a simpler algorithm becomes available: we can immediately compute
the toy angle-action variables of $\vw$ and then adjust $\vJ$ until the point
$\vw'(\vthetaT,\vJ)$ pointed to by the torus $\vJ$ at the toy angles of $\vw$
coincides with $\vw$. In practice the code computes the toy actions
$\vJT'(\vthetaT,\vJ)$ assigned by $\vJ$ to $\vw'$ and adjusts
$\vJ$ until $\vJT'$ and $\vJT$ coincide. In the axisymmetric case,  the
St\"ackel Fudge returns the correct value of $J_\phi$, so the search is
two-dimensional.

The torus returned by {\tt ActionFinderTG} for actions $\vJ$ may require more
terms $S_\vk$ in its generating fuction, than the torus returned by {\tt
fitTorus} for $\vJ$ because {\tt ActionFinderTG} doesn't optimise the toy map
for $\vJ$. This shortcoming of the returned torus does not significantly
impact the returned angle-action coordinates and a fully optimised torus can be
obtained by calling {\tt fitTorus} on $\vJ$.
}

\section{What's in the toolbox}\label{sec:inBox}

We now illustrate the use of the new tools with snippets of C++ code --
Python analogues could be written. First
we declare a potential, which could be the potential of a sophisticated
Galaxy model such as that of \cite{BinneyVasiliev2023}, but for simplicity we
adopt the potential generated by an oblate double-power-law mass
distribution that's essentially a flattened \cite{He90} model.
 {\obeylines\tt\parindent=10pt 
potential::PtrPotential pot = 
potential::createPotential(utils::KeyValueMap(
"type=spheroid, gamma=1, beta=4, scaleradius=1, 
q=0.6"));
}

\subsection{The class {\it TorusGenerator}}

\noindent Tori are created by an
instance of the class {\it TorusGenerator}:
{\obeylines\tt\parindent=10pt
double tol=5e-5;
	TorusGenerator TG(*pot, tol);
}
\noindent The second argument of the creator is
an indication of how small we require the dispersion in $H$ to be around a
torus: the Levenberg-Marquardt algorithm exits when
{\obeylines\tt\parindent=10pt
Hdisp < tol*freqScale*Jscale
}
\noindent where {\tt Jscale} is $J_r+J_z$  and {\tt freqScale} is the circular frequency
at the radius $\Rsh$ of the shell orbit -- the orbit with the given
values of $J_z$ and $J_\phi$ but $J_r=0$.
{\tt TG} can now be used to generate any number of tori in the given
potential: the lines
{\obeylines\tt\parindent=10pt 
	Actions J(.1,.5,1);
	Torus T(TG.fitTorus(J));
}
\noindent will make {\tt T} a torus with $J_r=0.1$, $J_z=0.5$ and
$J_\phi=1$.  

If diagnostic information is required, a file name can be given
as a third argument of  {\tt TorusGenerator}.

\subsection{The class {\it Torus}}

We can
now discover the energy of this torus and its frequencies
{\obeylines\tt\parindent=10pt 
printf("E = \%f, Omegar = \%f$\backslash$n",
T.E, T.freqs.Omegar);
}
\noindent To obtain the coordinates of a point on the torus we write
{\obeylines\tt\parindent=10pt 
	Angles theta(1.5, 3.2, 0);
	coord::PosMomCyl RpR(T.from\_true(theta));
	coord::PosVelCyl RvR(toPosVelCyl(RpR));
}
\noindent 
 {\tt T}'s method {\tt from\_true} returns the data type {\tt
coord::PosMomCyl} which comprises the position in cylindrical polars and the
canonically conjugate momenta $(p_R,p_z,p_\phi)$. This is a new data type
within \agamaTwo\ introduced to facilitate working with canonical
transformations. In the case of cylindrical coordinates $p_R=v_R$ and
$p_z=v_z$ so the only difference from the older data type {\tt
coord::PosVelCyl} is that $p_\phi=Rv_\phi$. In the spherical case the
difference between {\tt PosMomSph} and {\tt PosVelSph} is larger: $p_r=v_r$
but $p_\vartheta=rv_\vartheta$ and $p_\phi=r\sin\vartheta v_\phi$.  The
method {\tt from\_true} returns the phase-space point corresponding to the
true angle variables {\tt theta}; {\tt T} has a method {\tt from\_toy} to
return the location corresponding to toy angles, $\vthetaT$.

\begin{figure}
\centerline{\includegraphics[width=.8\hsize]{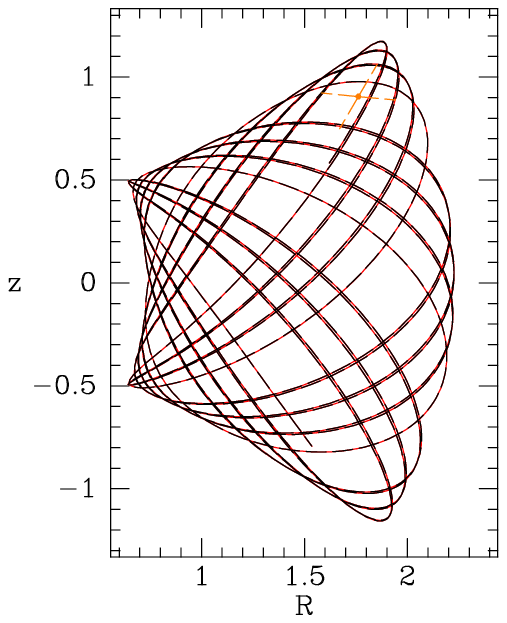}}
\caption{In black an orbit produced by the {\it orbit} method of a torus and
in red the result of directly integrating the equations of motion. The orange
cross at upper right
shows the velocities with which the torus will eventually visit the
marked point, according to the method {\it containsPoint}. The orbit's
actions are $(J_r,J_z,J_\phi)=(0.125,0.125,0.5)$ and its energy is
$E=-0.2878$.}\label{fig:t-seqs}
\end{figure}

\noindent We can display the orbit {\tt T} describes by writing
{\obeylines\tt\parindent=10pt 
	double duration=4000, dt=2;
	std::vector<std::pair<coord::PosVelCyl,double> > 
	traj(T.orbit(theta, dt, duration));
}
\noindent
After these lines have been executed the vector  {\tt traj} will contain in
its {\tt first} attribute 2000 phase-space positions  separated by 2 time units along the
orbit starting from the location pointed to by the true angles {\tt theta}.
The black curve in Fig.~\ref{fig:t-seqs} joins the $R$ and $z$ coordinates
obtained in this way. The black curve essentially obliterates a  red curve
that was  obtained by integrating the equations
of motion directly:
{\obeylines\tt\parindent=10pt 
	traj.clear();
	traj=orbit::integrateTraj(RvR,duration,dt,*pot);
}
\noindent because at the precision of the plot the agreement between the two time
series is exact.

The method {\it containsPoint} determines whether a given spatial location
$\vx$ is ever visited by {\tt T}, and if so at what values of angles and
velocities. The orange cross  at upper right of Fig.~\ref{fig:t-seqs} shows the
velocities at which {\it containsPoint} predicts the orbit will eventually
reach the marked point. These were obtained by writing
{\obeylines\tt\parindent=10pt 
	std::vector<Angles> angs;
	std::vector<coord::VelCyl> vels;
	std::vector<double> Jacobs;
	if(!T.containsPoint(w,angs,vels,Jacobs,1e-6))
		printf("Point not found$\backslash$n");
}
\noindent after which {\tt angs} contains the four true angles of visits to
{\tt w},
{\tt vels} contains
the corresponding velocities and {\tt Jacobs} contains the Jacobians whose
inverses are the contributions to the density of the orbit at the given
location {\tt w}. The last argument of {\it containsPoint} determines the
precision required in {\tt theta}.

\begin{figure}
\centerline{\includegraphics[width=.8\hsize]{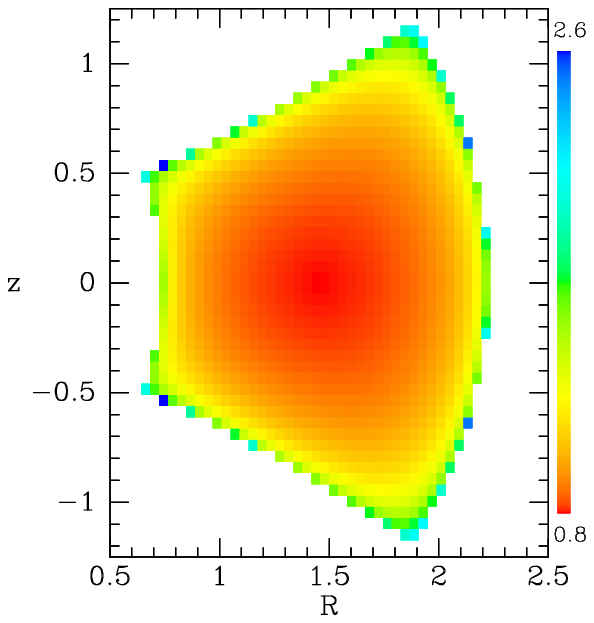}}
\centerline{\includegraphics[width=.8\hsize]{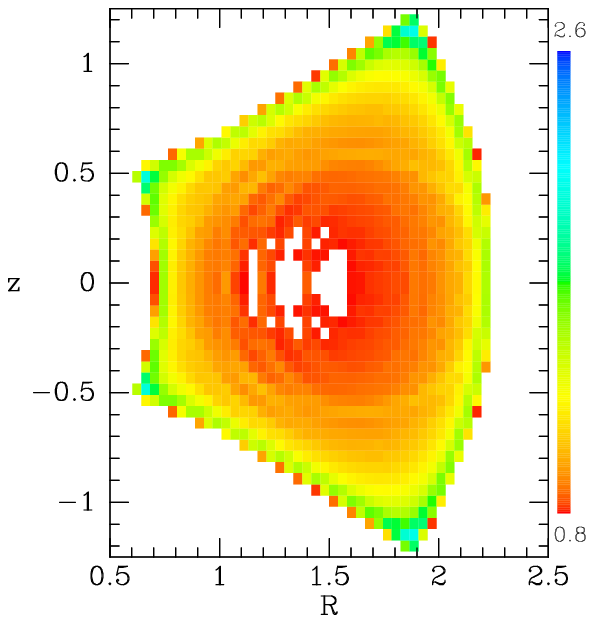}}
\caption{Plots of the logarithm of the density contributed by stars on the
torus of Fig.~\ref{fig:t-seqs}. The upper plot was made using the method {\it density} while the lower
plot was made by uniformly sampling angle space at four times as many points
as there are pixels in each panel.}\label{fig:densities}
\end{figure}

The density $\rho(\vx)$ contributed at $\vx$ by the torus with actions $\vJ$ can be determined from the DF
$f(\vJ')=\delta(\vJ'-\vJ)$ and writing
\begin{align}\label{eq:rho}
\rho(\vx)&=\int\d^3\vv\,f[\vJ'(\vx,\vv)]\cr
&=\int\d^3\vJ'\,{\p(\vv)\over\p(\vJ')}\delta(\vJ'-\vJ)={\p(\vv)\over\p(\vJ)}.
\end{align}
The Jacobian between any two systems of canonical coordinates is unity; in
particular
\[\label{eq:Jacob}
\d^3\vtheta\,\d^3\vJ=\d^3\vx\,\d^3\vv
\]
Dividing through by $\d^3\vJ\,\d^3\vx$ it follows that
\[
\Big({\p(\vv)\over\p(\vJ)}\Big)_\vtheta=\Big({\p(\vtheta)\over\p(\vx)}\Big)_\vv
\]
 Using this relation in equation (\ref{eq:rho}) we have that the density
contributed by {\tt T} to $\vx$ at any visit is $\p(\vtheta)/\p(\vx)$. The
method {\it containsPoint}
returns in {\tt Jacobs} the inverses of these densities; it returns the
inverses because these tend to zero as the point approaches the edge of the
region visited by {\tt T}, so the density diverges there. The method {\it
density} returns the density itself: the lines
{\obeylines\tt\parindent=10pt 
coord::PosCyl Rz(1.5, 1.4, 0);
double rho=T.density(Rz);
}
\noindent
leave in $\tt rho$ the density at the specified point contributed by all
visits. The upper panel of Fig.~\ref{fig:densities} was made by calling {\it
density} for each cell of a $50\times50$ grid in $(R,z)$. The lower panel was
obtained by uniformly sampling a $100\times100$ grid in angle space and
adding one unit of mass to the plotted $50\times50$ grid in real space at the
location {\it from\_true} returned for that angle. We see that calling {\it
density} produces a significantly smoother plot -- in the lower plot a good number of cells in
the low-density inner region are white because they are  not reached from any of the
$10^4$ locations in angle space. Integrating the orbit and
accumulating mass over $10\,000$ timesteps would have produced an even less
smooth plot, so {\it containsPoint} promises to be useful in Schwarzschild
modelling.

\subsection{Surfaces of section}

\begin{figure}
\centerline{\includegraphics[width=\hsize]{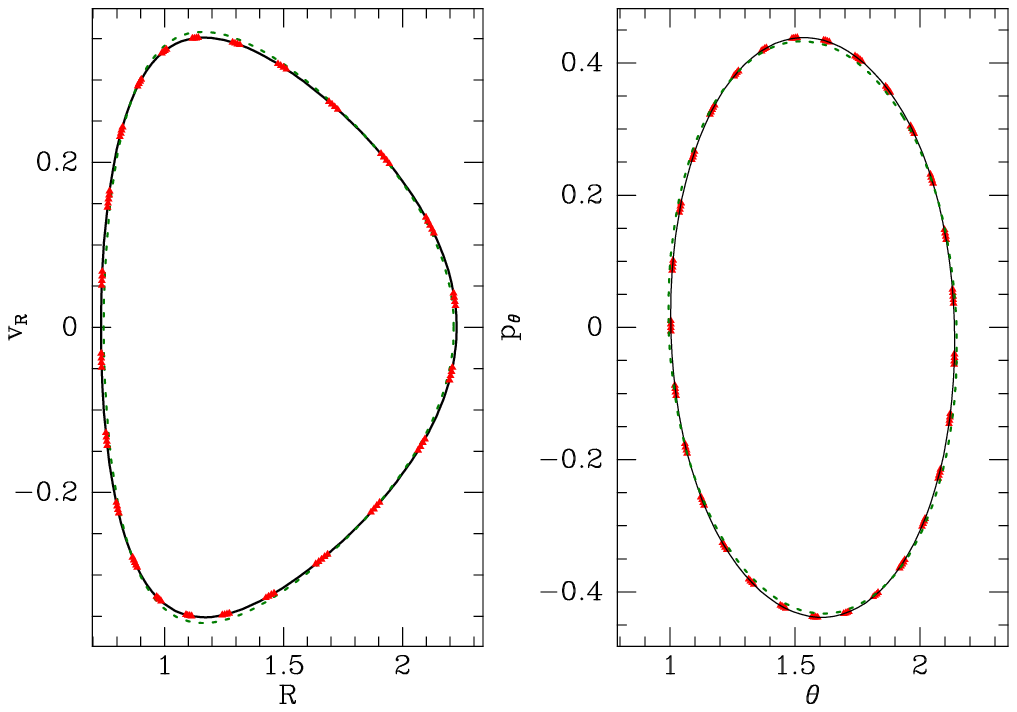}}
\caption{$(R,p_R)$ (left) and $(\vartheta,p_\vartheta)$ surfaces of section
for the  orbit plotted in Fig.~\ref{fig:t-seqs}.
The green curves are provided by a toy
map that comprises the isochrone orbit of the given actions mapped into the
cylindrical system by a point transformation defined by a choice of $\Delta$
and the functional relation $v(\vartheta)$ shown in
Fig.~\ref{fig:thetaPsi}}\label{fig:Xsecs}
\end{figure}

Poincar\'e surfaces of section are invaluable diagnostics of the structure of
phase space. A torus yields a curve in a surface of section that is
essentially a cross section of the torus.  The method {\it zSoS} produces
this curve: the black curve in the left panel of Fig.~\ref{fig:Xsecs} was
produced by executing 
{\obeylines\tt\parindent=10pt 
 std::vector<double> Rs, vRs;
 T.zSoS(Rs,vRs,Npt,Rmin,Rmax,Vmax);
}
\noindent After execution of these
statements, the vectors {\tt Rs} and {\tt vRs} contain the values of $R$ and
$v_R$ at which $z=0$ with $v_z>0$, while the doubles {\tt Rmin} and {\tt Rmax} contain the smallest
and largest values of $R$ on the curve and {\tt Vmax} contains the largest
value of $v_R$ to facilitate plotting. The small red triangles in the left
panel of  Fig.~\ref{fig:Xsecs} are consequents obtained by integrating the equations of motion from one
point on the torus -- they were obtained by executing
{\obeylines\tt\parindent=10pt 
 orbit::makeSoS(Rzv, *pot, Rs, vRs, Rbar, 
thetas, pthetas, 100);
}
\noindent The dashed green curve in Fig.~\ref{fig:Xsecs} is the surface of section of the
torus that comprises the toy map without a generating function
$S(\vJ,\vthetaT)$. It is barely distinguishable from the black curve along
which the red consequents lie. This is an example of the precision with which
the new, very flexible point transformations can fit orbital tori even in the
absence of a generating function $S(\vthetaT,\vJ)$.

The right panel of Fig.~\ref{fig:Xsecs} shows the $(\vartheta,p_\vartheta)$
surface of section of the same orbit. A dashed black curve of 100 points
joins sets of small red triangles. The black curve was created by
executing 
{\obeylines\tt\parindent=10pt 
	std::vector<double> thetas, pthetas;
 	T.rSoS(thetas,pthetas,Rbar,100,tmax,ptmax); 
} 
 \noindent which fills the vectors {\tt thetas} and {\tt pthetas} with the
$(\vartheta,p_\vartheta)$ values at which the orbit crosses $r={\tt Rbar}$ with
$v_R>0$. The last two arguments of the call give the largest encountered values of
$\vartheta$ and $p_\vartheta$. Data for the triangles were produced by the
statement {\tt orbit::makeSoS} above. As in the left panel, a dashed green
curve shows the torus that's provided by the toy map alone.

\begin{figure}
\centerline{\includegraphics[width=\hsize]{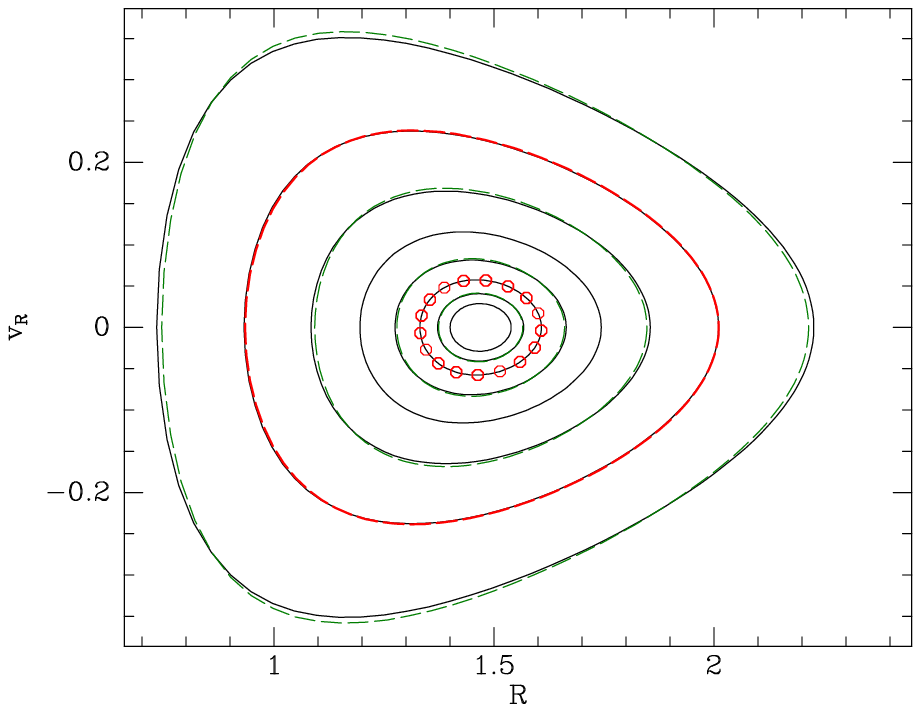}} \caption{A
surface of section at energy $E=-0.2878$ constructed by using the method {\it
Torus::zSoS} on the sequence of tori returned by the method {\it
TorusGenerator::constE}. The black curves were produced by tori in that
sequence with values of $J_r$ that decrease from $0.125$ to $0.001$; the
broken red curve was produced by a torus interpolated between the first and
third tori in the sequence for the radial action of the second torus in the
sequence. The red circles mark consequents produced by Runge-Kutta integration
of the equations of motion from an initial condition provided by the torus
with third smallest value of $J_r$. The broken green curves were produced by
the toy maps of alternate tori acting alone.  }\label{fig:logSoS}
\end{figure}

An excellent way to identify the impact of resonances is to integrate several
orbits at a common energy and plot their  $(R,v_R)$   consequents together.
Since the energy of a torus is unknown until after it has
been constructed, {\it TorusConstructor} has a method {\it constE} that
constructs a series of tori of a common energy and angular momentum $J_\phi$
but decreasing values of $J_r$ and increasing ones of $J_z$.
The tori that contribute the black curves to Fig.~\ref{fig:logSoS} were constructed by the
lines
 {\obeylines\tt\parindent=10pt 
	double Jrmin = 0.001;
	std::vector<Torus> Ts = TG.constE(Jrmin,J,7);
}
\noindent The tori are logarithmically spaced in $J_r$ between the values of
$J_r$ of the starting actions $\vJ$ and the minimum radial action {\tt Jrmin}. The
broken green curves show surfaces of section produced by the toy maps of
alternate tori as an indication of what has to be done by the generating
function $S(\vthetaT,\vJ)$ -- this vanishes with $J_r$ so the smallest green
curve almost obliterates its black partner.

The red curve in Fig.~\ref{fig:logSoS} was produced by a torus interpolated
for the action $J_r$ of the second torus between the first and third tori in
the sequence returned by {\it constE}. The relevant lines are
 {\obeylines\tt\parindent=10pt 
	double dJ = (Ts[1].J.Jr-Ts[0].J.Jr);
	double DJ = (Ts[2].J.Jr-Ts[0].J.Jr);
	Torus T(interpTorus(dJ/DJ,Ts[2],Ts[0]));
	T.zSoS(Rs,vRs,100,rmin,rmax,VRmax);
}
\noindent The closeness with which the red curve overlies the black curve
demonstrates that by interpolation one can create  additional tori extremely
cheaply.

All tori contributing to Fig.~\ref{fig:logSoS} have isochrone toy maps. Since
interpolate between tori that use different toy map types is impossible,
when {\tt InterpTorus} is required
to interpolate between tori with different toy map types, it starts by creating a
torus that uses harmonic-oscillator maps at the actions of the torus that
uses isochrone maps. That done it  interpolates.

\begin{figure}
\centerline{\includegraphics[width=.8\hsize]{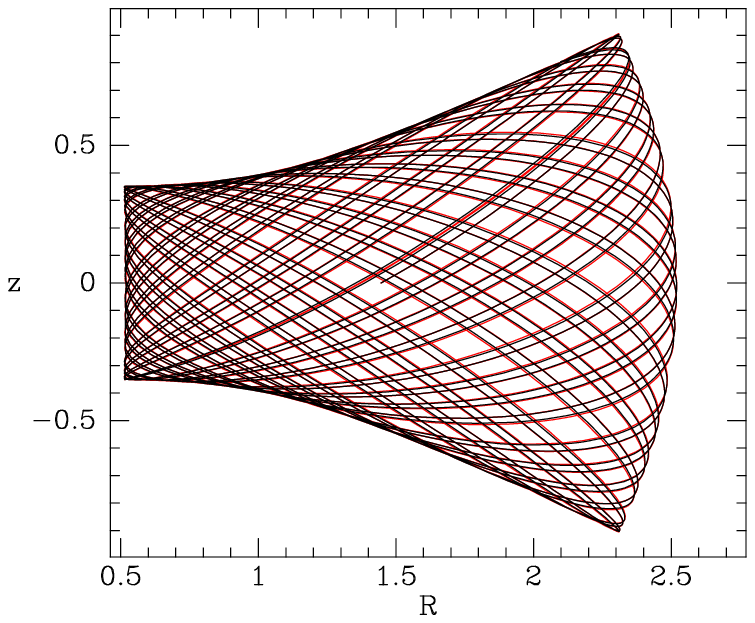}}
\centerline{\includegraphics[width=.8\hsize]{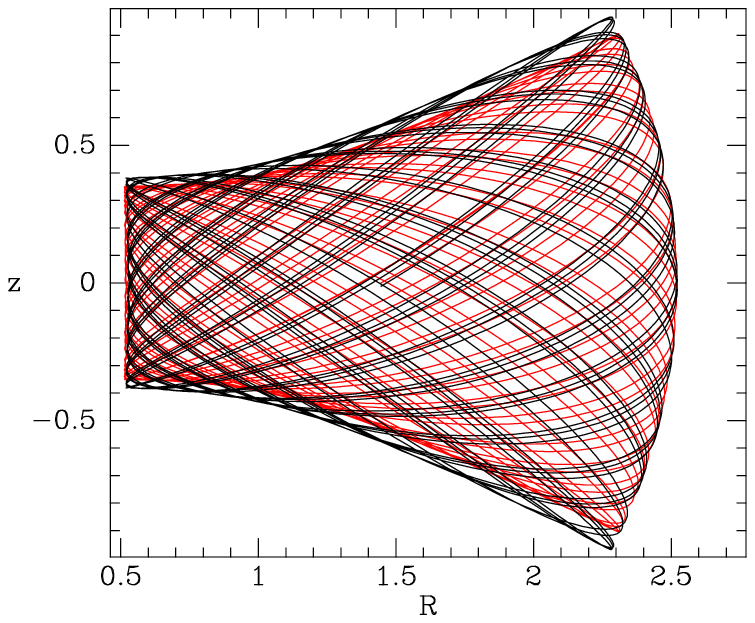}}
\caption{Testing the new action finder. In both panels an eccentric orbit
in an NFW potential is plotted in red and then over-plotted in black by a
time sequence obtained from a torus, with the location at $t=0$ taken to be
the orbit's initial condition. The tori for the upper and lower panels were
obtained by applying to the orbit's initial condition the new action finder
and the standard St\"ackel-Fudge action finder, respectively.
}\label{fig:AFs}
\end{figure}

\subsection{The action finder}\label{sec:afTsec}

{\rd
We have added to \agamaTwo\ an {\it ActionFinder} in
which} a {\it TorusGenerator} does the work of computing $(\vtheta,\vJ)$ from
$(\vx,\vv)$. The line
 {\obeylines\tt\parindent=10pt 
	ActionFinderTG afT(pot, TG);
}
\noindent creates an object {\tt afT} that has the same methods as the standard
St\"ackel Fudge {\it ActionFinder}, so given a phase-space location in cylindrical coordinates
{\tt w}, the corresponding angle-action coordinates  can be obtained from
the line
 {\obeylines\tt\parindent=10pt 
	ActionAngles aa(afT.actionAngles(w));
}
\noindent Since {\it afT} determines the torus through the given
point in addition to determining the point's angle-action variables, {\tt
actionFinderTG} has an additional method that returns the torus free of charge:
\noindent
 {\obeylines\tt\parindent=10pt 
	Torus T;
	ActionAngles aa(afT.actionAnglesTorus(w,T));
}

\begin{figure} 
\centerline{\includegraphics[width=.8\hsize]{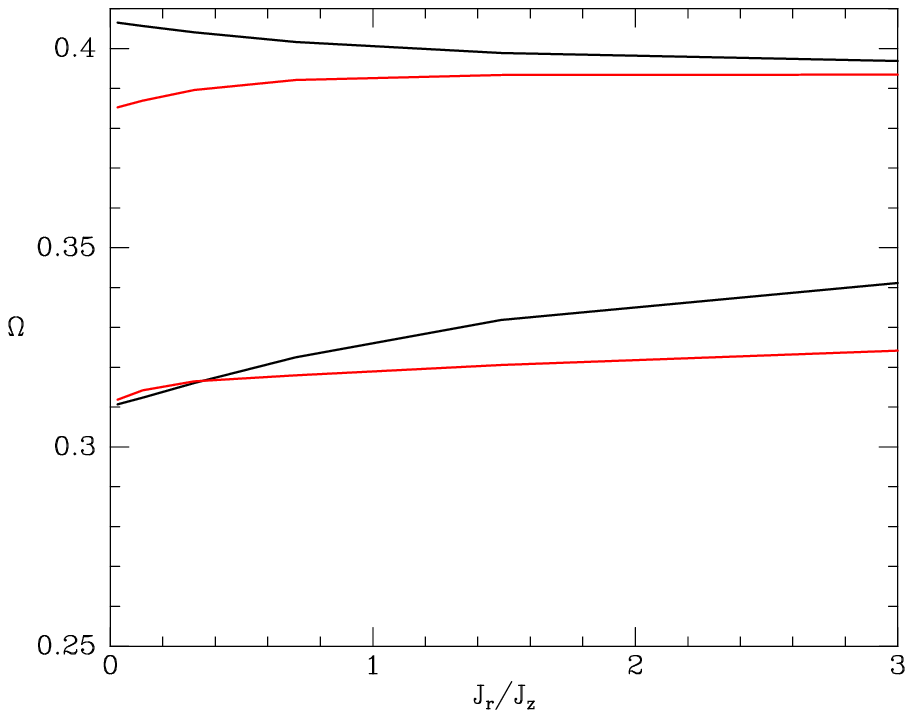}}
\caption{The frequencies returned by the two action finders: values of
$\Omega_r$ (lower curves) and $\Omega_z$ (upper curves) returned by the Fudge
are plotted in red, while those from the new action finder are plotted in
black. The horizontal variable is $J_r/J_z$ for a sequence of orbits at
$E=-0.3$. }\label{fig:Omegas}
\end{figure}

Fig.~\ref{fig:AFs} demonstrates how much better the new action finder
performs than that based on the St\"ackel Fudge by plotting orbits in the
NFW-like potential that \cite{WrightB} found problematic. This
potential is created
by the statement
\noindent
 {\obeylines\tt\parindent=10pt 
	pot=potential::createPotential(
	utils::KeyValueMap("type=spheroid gamma=1 beta=3 
	outercutoffradius=10 axisratioz=0.5"));
}
\noindent
In Fig.~\ref{fig:AFs}  black lines showing the time
series produced by a torus are plotted on top of the orbit computed by Runge-Kutta and
drawn in red. In the upper panel the torus was returned by {\it
ActionFinderTG} from the orbit's initial condition and its time series
essentially obliterates that from the Runge-Kutta because the new action
finder has inferred the correct torus from the initial condition. In the
lower panel the black time series is clearly distinguishable from the red
numerically integrated orbit on account of significant errors in the actions
derived by the Fudge from the initial condition. Quantitatively, the Fudge
returns $J_z=0.0670$ while the correct value is $J_z=0.0582$.

Frequencies returned by the Fudge are particularly suspect
\citep{WrightB}.  In Fig.~\ref{fig:Omegas} $\Omega_r$ and $\Omega_z$
are plotted against $J_r/J_z$ for a sequence of orbits at the same energy and
increasing eccentricity. The black curves show values returned by the Fudge
while red curves show the frequencies of the  returned tori. The coincidence
of the black and red lines in the upper panel of Fig.~\ref{fig:AFs} attests to
the validity of the frequencies of tori, so the significant divergence of the
red and black curves in Fig.~\ref{fig:Omegas} indicates that the frequencies
returned by the Fudge are systematically in error.

\subsection{The class {\it eTorus}}

Real galactic potentials are not integrable: they do not admit a global set
of angle-action coordinates. As discussed in \cite{Binney2016}, if one
sets a very tight constraint {\tt tol} on the permitted dispersion in $H$,
the tori constructed for adjacent actions may cross.
This unphysical phenomenon arises because the torus mapper is attempting to
generate a global set of angle-action coordinates when no such set is
possible. As Binney explains, in these circumstances one should set an
undemanding upper limit on $\sigma_H$ such that tori never intersect. Then
the set of tori comprise a global system of angle-action coordinates for a
Hamiltonian $H_0(\vJ)$ that differs slightly from the real one. The difference between
the real and true Hamiltonians  can be quantified by Fourier analysing the
real Hamiltonian over the tori of the constructed Hamiltonian. That is, one
computes the quantities $h_\vk$ that are defined by
\[\label{eq:defhn}
H(\vtheta,\vJ)=H_0(\vJ)+\sum_{\vk\ne0} h_\vk(\vJ)\, \e^{\i(\vk\cdot\vtheta)}
\] 
Typically $h_\vk$ will be much smaller than $H_0$ and it can be safely
neglected except where $\Omega_{\rm slow}=\vk\cdot\vOmega$ is small. Where
$\Omega_{\rm slow}$ is small, resonant trapping may occur.  Its effects can
be computed from $h_\vk$ with Hamiltonian perturbation theory.

Hamiltonian perturbation theory can also be used to study the dynamics of
potentials for which we cannot (currently) do torus mapping directly. For
example, the dynamics of rotating bars can be studied by treating the non-axisymmetric
part of a bar's potential as a perturbation on an axisymmetric potential
\citep{Binney2018,Binney_negJ}.

An {\it eTorus} combines a torus of the  integrable Hamiltonian $H_0$ with
the Fourier coefficients $h_\vk$ -- the latter can be used to create tori of
resonantly trapped orbits by perturbation theory.
To create an {\it eTorus} we
write
 {\obeylines\tt\parindent=10pt 
potential::bar\_potS bar(0.4, 220./978., 1.7);
eTorus eT(TG.fiteTorus(J,\&bar));
}
\noindent Here {\tt bar} is the non-axisymmetric part of a bar's potential:
in this example it takes the form $\Phi_2(R,z)\cos(2\phi)$. The resulting
Fourier coefficient will be dominated by ones generated by the bar. If the
{\tt bar} argument of the {\it eTorus} creator were omitted, the Fourier
coefficient would be smaller and any resonant trapping that they described
would be inherent to the axisymmetric potential.

\begin{figure}
\centerline{\includegraphics[width=.9\hsize]{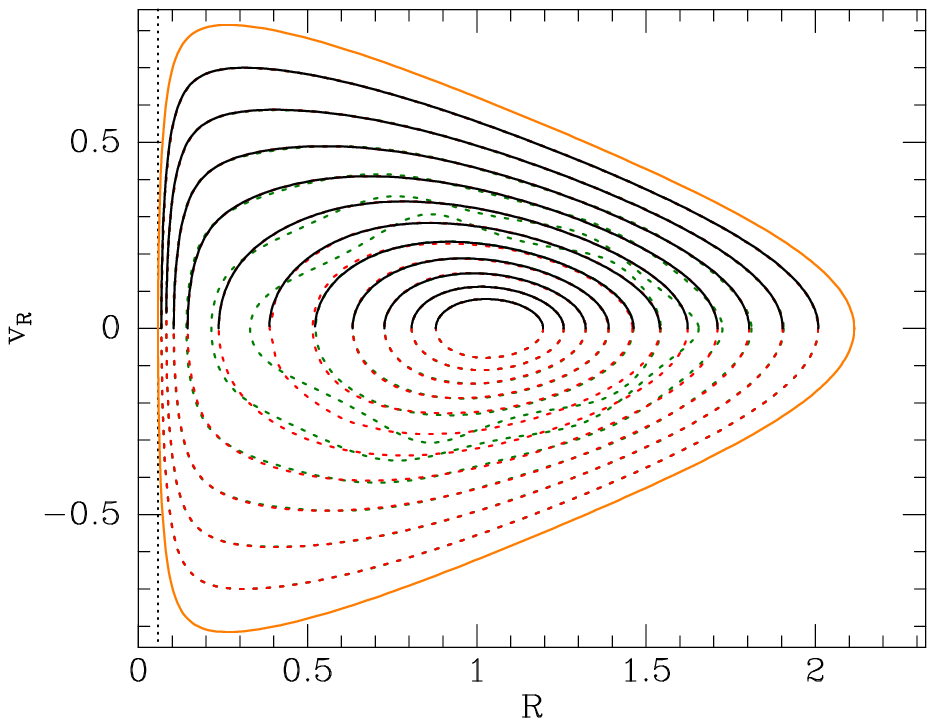}}
 \caption{$(R,v_R)$ surface of section for a perfect ellipsoid with unit mass
and scalelength and axis
ratio $c/a=0.6$ at energy $E=\Phi(0,0)/2$ and $J_\phi=0.05$.  Black and
orange curves are obtained from St\"ackel's analytic formulae -- the orange
curve is a cross section through the torus $J_z=0$, which has the largest
value of $J_r$ obtainable at this $(E,J_\phi)$ combination. The red dashed curves
were produced by a {\it TorusGenerator}; at $v_R>0$ the red curves are
largely obliterated by the over-plotted black curves. The tori that drew the
six outer red curves used harmonic oscillators, while the five inner
red curves used isochrones. The dashed green curves show cross sections
through the tori provided by the toy maps alone. The largest black curve is
for $(J_r,J_z)=(0.304,0.072)$ while the smallest black curve is for
$(0.0062,0.529)$.}\label{fig:StackSoS}
\end{figure}

The $h_\vk$ are wrapped into an instance of the class {\it
PerturbingHamiltonian}, which comprises two {\tt std::vector}s: {\tt
indices}, containing the $\vk$ vectors of the coefficients, and {\tt values},
containing the complex coefficients themselves. Perturbing Hamiltonians can
be interpolated so {\it eTori} can be interpolated.

\subsection{Coverage of phase space}

The introduction of harmonic-oscillator HJ maps and more flexible point
transformations has materially extended the range of tori that can be
successfully generated. In particular, tori can now be constructed for highly
eccentric orbits that the BM16 mapper failed on.  Fig.~\ref{fig:StackSoS}
illustrates this fact by plotting a surface of section for a perfect
ellipsoid \citep{deZeeuw1985} at a small value ($0.05)$ of $J_\phi$. The
black curves are obtained from St\"ackel's analytic formulae, while the red
and green curves were produced by a {\it TorusGenerator}: the red curves
include the generating function (\ref{eq:defS}) while the green curves use
only the toy map. The six largest black curves were generated by
harmonic-oscillator maps while the five smaller curves were generated by
isochrone maps. See
Section~\ref{sec:ToyMap} for an explanation of when and why a harmonic
oscillator provides a better HJ map than an isochrone. The BM16 mapper, upon
which the torus mapper of \agama\ is based, is unable to generate most of
these tori. It abandons construction because getting the \rms\ in $H$ down
requires an increase in the number of $S_\vk$ employed, but adding $S_\vk$
increases the frequency at which negative toy actions are encountered. The
key to escaping this bind is to upgrade the toy map.

{\rd
We set both old and new torus mappers to create tori for the perfect
ellipsoid with unit mass and scalelength and axis ratio $c/a=0.6$. The tori
formed a $4^3$ grid in action space: each action set in turn to
$0.01,\,0,05,\,0.25$ and 1. With $J_\phi=0.01$ the old mapper returned a
torus in only 3/16 cases, namely when $J_r\ge0.05$ and $J_z=1$.  With
$J_\phi=0.05$ the old mapper returned a torus in 8/16 cases, always failing
when $J_r=1$ and sometimes when $J_r=0.25$ or $0.01$. With $J_\phi=0.25$, the
old mapper failed when $J_r=1$ but otherwise returned a torus. With
$J_\phi=1$, the old mapper returned a torus except when $J_r=1$ and
$J_z\ge0.25$. The new mapper produced tori in every case although in one
case, $\vJ=(1,1,.05)$, sections of the $(R,z)$ plot of the orbit were marred
by wiggles arising from poor convergence of the series for the generating
function $S(\vJ,\thetaT)$. Increasing the parameters {\tt timesr} and {\tt
timesz} in the routine {\tt fitTorus} suppressed the wiggles.
}

\section{Application to stellar streams}\label{sec:streams}

The advent of data from Gaia has led to the detection of stellar streams in
industrial quantities \citep[e.g.][]{Ibata2024}. It has long been recognised
that streams are best quantified by angle-action coordinates
\citep{Tr99,HeWh99}, and that those seen at high Galactic latitudes are
potentially powerful probes of our Galaxy's gravitational potential
$\Phi(\vx)$ and
hence its dark-matter distribution
\citep[e.g.][]{Ibea01,JJBstream2008,EyJJB11,Sa14,Erkal2019,Ibata2024}.

Even in the era of Gaia, the phase-space locations of individual stars have
significant uncertainties so we must constrain $\Phi$ by computing
the probability distribution for the density of stars in some image of
phase space as functions of parameters describing the stream, and then
marginalise the likelihood of the observed distribution of stars over the
stream parameters \citep{Bovy2014:streams}. To implement this agenda we need
an algorithm that rapidly generates model streams given a potential and
stream parameters. We now show how the {\it TorusGenerator} can be used to
implement a slight modification of the algorithm described by
\cite{Bovy2014:streams}. 

The demonstration will serve to illustrate facilities within \agamaTwo\ for
connecting astrophysical structures to observational data, which typically
involve sky coordinates, proper motions, magnitudes, etc.  Functions for
connecting models to observational data are gathered into \agamaTwo's {\tt obs}
namespace and listed in Appendix \ref{sec:obs}.

\subsection{Generating a stream}

We start by choosing the actions {\tt JP} of the stream's progenitor,
computing its torus {\tt TP} and finding  the times {\tt tperi} of ten 
pericentres
 {\obeylines\tt\parindent=10pt 
	Actions JP(288.5*h, 897.6*h, 3173.7*h);
	Torus TP(TG.fitTorus(JP));
	double TPr = 2*M\_PI/TP.freqs.Omegar;
	int Np=10; std::vector<double> tperi(Np);
	for(int i=0; i<Np; i++) tperi[i]=i*TPr;
}\noindent
Here {\tt h=intUnits.from\_Kpc\_kms} is \agamaTwo's unit of action and the
values of the progenitor's actions are those used by \cite{Bovy2014:streams}
to facilitate comparison with that paper. We do however take $J_\phi=+3173h$
rather than $-3173h$ because \agamaTwo\ follows the {\sc astropy} convention
in using a right-handed coordinate system such that the Sun's value of
$J_\phi<0$. Bovy used the standard logarithmic potential with circular speed
$V_c=220\kms$. This potential cannot be used with \agamaTwo's action finders
because they assume that $\Phi(\vx)\to0$ as $|\vx|\to\infty$.  A potential
that is functionally equivalent to the standard one but tends to zero at
infinity is
\[\label{eq:flatVc}
\Phi(\vx)=-\fracj12V_c^2{L_c\over1+L/L_c}
\]
where $L=\ln m^2$ with $m^2=1+(R^2+z^2/q^2)/R_c^2$ and $L_c$ a suitably large constant:
we used 1000 by executing the line
 {\obeylines\tt\parindent=10pt 
	potential::PtrPotential pot(new 
	\qquad	potential::Logarithmic(Vc,Rc,1,ZtoX,1000));
}
\noindent with {\tt Rc = 0.1*intUnits.from\_Kpc}.

\begin{figure}
\centerline{\includegraphics[width=\hsize]{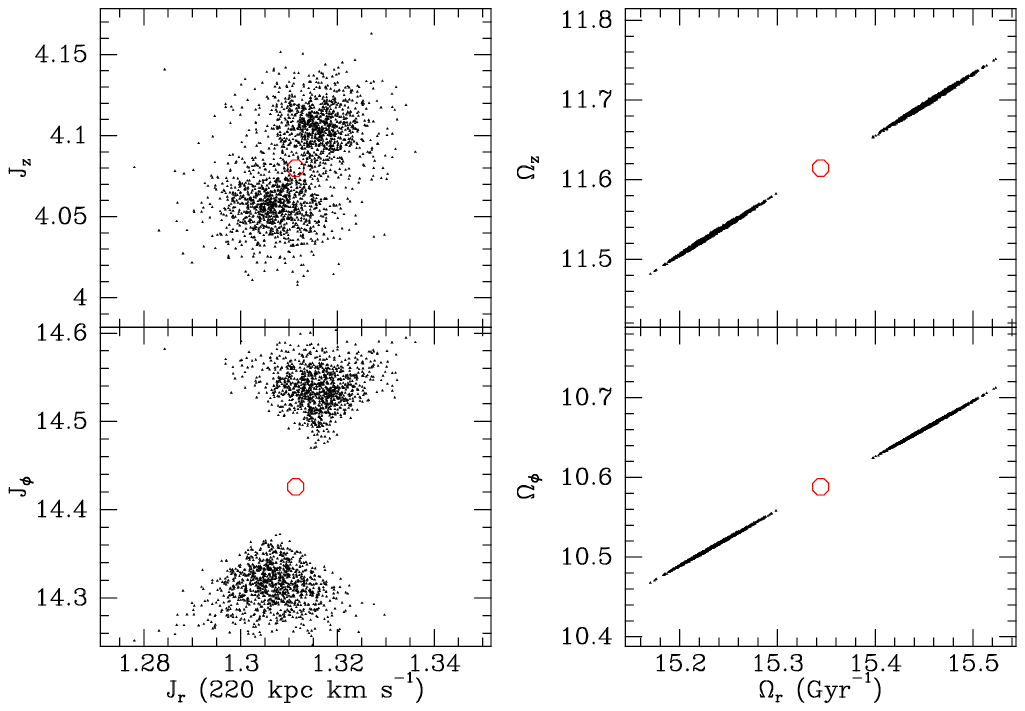}}
\caption{The distribution of 2000 stars in  action and frequency space. The
plot was made by interpolating for each of the leading and trailing streams
1000 tori from grids that covered the regions occupied by the streams. The
plot may be compared with Fig.~2 of Bovy (2014), which shows data from a full
N-body simulation. The data for these plots required 3 seconds to compute on
a laptop.} \label{fig:JJfig}
\end{figure}

Stars removed from the cluster form two clouds in action space, one below and
one above the progenitor's value of $J_\phi$ \citep{EyJJB11}. In real space
the lower cloud forms the leading tidal arm while the upper cloud forms the
trailing arm \citep{EyJJB11,Bovy2014:streams}. The clouds are triangular in
shape. We model this shape by distributing $J_r$ and $J_z$ normally around
mean values that are slightly offset from the corresponding actions of the
progenitor with dispersions that increase with the action-space distance
$|J_\phi-JP_\phi|$ from the progenitor. $J_\phi$
is also normally distributed and all three dispersions in action are related
to the progenitor's velocity dispersion by \citep[cf.][eqs
5--7]{Bovy2014:streams}
  {\obeylines\tt\parindent=10pt 
	double sigCluster = 0.365*intUnits.from\_kms;
	double DJr = sigCluster*(RZapo.R-RZperi.R)/M\_PI;
	double DJz = sigCluster*2*RZzmax.z/M\_PI;
	double DJphi = sigCluster*RZperi.R;
}
\noindent where {\tt RZperi} and {\tt RZapo} are the locations given by the
torus at angles (0,0,0) and $(\pi,0,0)$, while {\tt RZmax} is the location at
angles $(\pi,\pi/2,0)$. We create grids of tori that cover the volumes
occupied by the leading and trailing clouds.  For the leading cloud:
  {\obeylines\tt\parindent=10pt 
	int nx=5, ny=5, nz=5;
	std::vector<Torus> Ts(nx*ny*nz);
	std::vector<double> xs(nx), ys(ny), zs(nz);
	Actions Joff(0.7*DJr, 1.5*DJz, -4*DJphi);
	for(int i=0; i<nx; i++)\{
		\quad xs[i]=JP.Jr+Joff.Jr+(i-nx/2)*DJr;
		\quad for(int j=0; j<ny; j++)\{
			\qquad ys[j]=JP.Jz+Joff.Jz+(j-ny/2)*DJz;
			\qquad for(int k=0; k<nz; k++)\{
				\quad\qquad zs[k]=JP.Jphi+Joff.Jphi+(k-nz/2)*DJphi;
				\quad\qquad Actions J(xs[i],ys[j],zs[k]);
				\quad\qquad Ts[(i*nx+j)*ny+k]=TG.fitTorus(J);
			\qquad\}
		\quad\}
	\}
	TorusGrid3 Tgrd(xs,ys,zs,Ts,TG);
}
\noindent From {\tt Tgrd} we can create the torus with actions {\tt J} by
linear interpolation at negligible
cost by the statement
   {\obeylines\tt\parindent=10pt 
	Torus T(Tgrd.T(J));
}
 \noindent In this way we create 1000 tori in each cloud by sampling the
Gaussianly distributed actions. Fig.~\ref{fig:JJfig} shows on the left two
projections of the resulting stellar stream in action space, and on the right
the corresponding projections in frequency space. The plot is very similar to
Fig.~2 of \cite{Bovy2014:streams}, which shows data from a full N-body
simulation. Filling out {\tt Tgrd} took three minutes on a laptop, but then a
stream with 2000 stars can be created in less than $10$ milliseconds. Note
that Bovy's figure relates to a stream with negative $J_\phi$ while
Fig~\ref{fig:JJfig} is for a stream with $J_\phi>0$ so the lower panels need
to be reflected in a horizontal plane before comparison with Bovy's plot. 

{\rd To constrain our Galaxy's dark-matter distribution, it's necessary to
consider streams in different potentials in addition to sampling the actions
of the progenitor. Tori can be created by interpolation for potentials that
lie between potentials for which tori have been constructed by the {\it
TorusGenerator}. Hence, there is no objection in principle to extending the
action-space grid used above to include the parameters of a family of
potentials. Then posterior probabilities could be assigned to potentials by
marginalising over progenitor actions.}

The near linearity of the frequency distributions in Fig.~\ref{fig:JJfig} is
a remarkable phenomenon given that these are the images of a near-spherical
cloud of objects in action space. As discussed by \cite{Tr99} and
\cite{EyJJB11}, it arises because for typical galactic potentials the
symmetric matrix ${\p\Omega_i/\p J_j}$ has one eigenvalue that is larger than
the others by a factor $20-30$, so a spherical region is mapped to an
ellipsoid that stretches along the corresponding eigenvector.

\begin{figure}
\centerline{\includegraphics[width=.8\hsize]{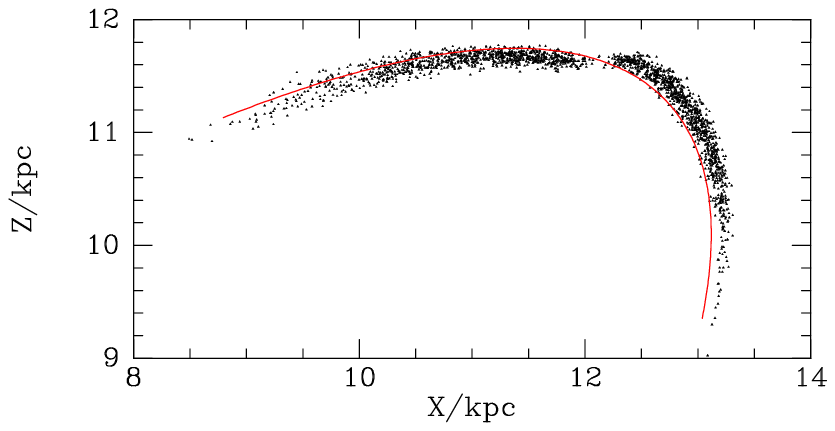}}
\centerline{\includegraphics[width=.8\hsize]{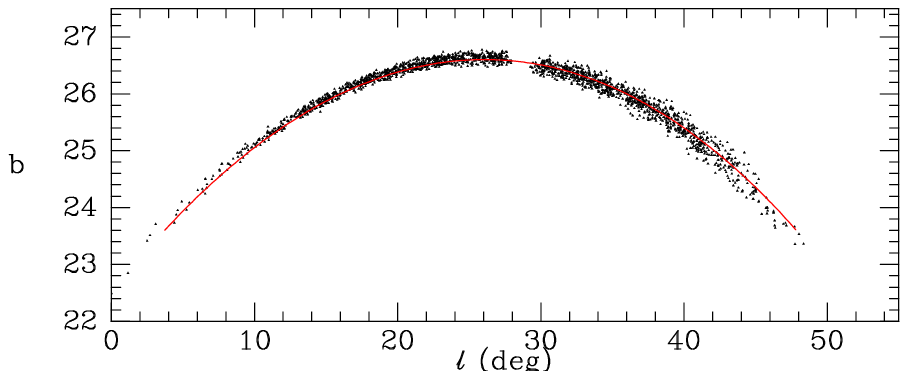}}
\caption{Model of the stream produced by the action-space distribution
plotted in Fig.~\ref{fig:JJfig}. The upper panel shows the $xz$ projection of
the stream, while the lower panel plots the stream on the sky in Galactic
coordinates. Dots represent the stream's 2000 stars and red curves show the
progenitor's orbit.}\label{fig:simpleStream0}
\end{figure}

To model the appearance of a stream on the sky, we must choose a distribution
of drift times $t_{\rm drft}$, that is the times that individual stars have
drifted free from the cluster. Bovy assumed a uniform distribution of $t_{\rm
drft}$ up to a maximum value $\sim5\Gyr$ associated with the cluster's
entering our Galaxy. Given that the Galaxy's tidal field is strongly
concentrated around pericentres \citep[e.g.][]{BinneyWarp2024}, it seems better
to model the distribution of $t_{\rm drft}$ by a sum of Gaussian distributions
of width $\tau\sim10\Myr$ around the time of each pericentre. With $t_{\rm
drft}$ chosen, a star's current angle variable is
\[
\vtheta=\vtheta_0+\Delta\vtheta+\vOmega_{\rm prog}(t-t_{\rm drft})+\vOmega t_{\rm drft},
\]
 where $\vtheta_0$ comprises the angle variables of the progenitor at $t=0$,
$\Delta\vtheta$ is the small change to $\vtheta$ as the star is extracted
from the cluster and $\vOmega_{\rm prog}$ and $\vOmega$ comprise,
respectively, the frequencies of the the progenitor and the star. {\rd We adopt
$\vtheta_0=0$.} Given $\vtheta$, the star's sky coordinates are easily
computed. Specifically, with {\tt T} a star's interpolated torus, the
required code is\footnote{The data types {\tt Actions}, {\tt Angles} and {\tt
Frequencies} can be multiplied by scalars and added, but this example
illustrates that {\tt Frequencies} must be cast to {\tt Angles} before
addition to {\tt Angles}.} 
   {\obeylines\tt\parindent=10pt 
Angles theta=theta0+dtheta
\qquad +(Angles)(TP.freqs*(t-td)+T.freqs*td);
coord::PosMomCyl qp(T.from\_true(theta));
coord::PosVelCyl Rzv(coord::toPosVelCyl(qp));
double s,Vlos;
PosVelSky lbmu(sun.toSky(Rzv,s,Vlos));
}
 \noindent After execution of these lines, {\tt lbmu.pos} contains the star's
current Galactic coordinates $(\ell,b)$, while the proper motion
$(\mu_\ell,\mu_b)$ comprise {\tt lbmu.pm}.  The variables {\tt s} and {\tt
Vlos} contain the star's distance in kiloparsecs and line-of-sight velocity
(in $\!\kms$).  Fig.~\ref{fig:simpleStream0} shows the stream produced by the
action-space distribution of Fig.~\ref{fig:JJfig} projected along the $y$
axis (upper panel) and projected onto the sky (lower panel). This may be
compared with Fig.~1 of \cite{Bovy2014:streams} or Fig.~3 of \cite{SaJJB13b}.
To achieve this good agreement the dispersion underlying $\Delta\vtheta$ has
to be small -- we adopted a Gaussian distribution with dispersion $0.001\pi$.

The object {\tt sun} used above to convert between Galacto-centric and
sky coordinates is an instance of the class {\tt obs::solarShifter}
and was created by executing
   {\obeylines\tt\parindent=10pt 
obs::solarShifter sun(intUnits);
}
\noindent Created thus the Sun's phase space Cartesian coordinates are taken to be
\[
(x,y,z,U,V,W)=(-8.27, 0, 0.025, 14, 251.3, 7)
\]
from \cite{Gravity2022}, \cite{ReidBrunthaler2020} and \cite{Schoenrich2012}. Another
location for the Sun can be specified by creating {\tt sun} with a second parameter that is a
pointer to an instance of {coord::PosVelCar} that contains the preferred
location.

\begin{figure}
\centerline{\includegraphics[width=\hsize]{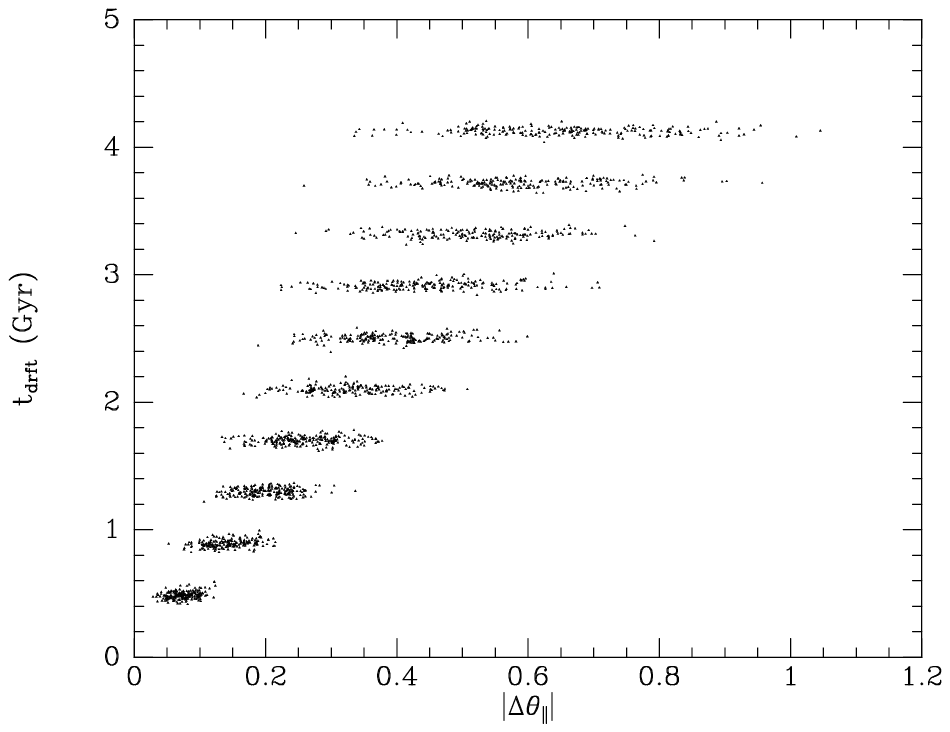}}
\caption{The distribution of $t_{\rm drft}$ versus
$|\Delta\theta_\parallel|$, the frequency difference projected along the
longest eigenvector of $\p\Omega_i/\p J_j$ between a star and the
progenitor.}\label{fig:simpleStream3}
\end{figure}

As we recalled above, a key role is played by the eigenvectors of the matrix
$\vM=\p\vOmega/\p\vJ$ evaluated at the progenitor's actions, {\tt JP}. This we compute by numerical
differencing -- the first row of $\vM$ contains derivatives w.r.t.\ $J_r$ so we
code
 {\obeylines\tt\parindent=10pt 
	double dJ=.001;
	Actions J(JP.Jr+.5*dJ, JP.Jz, JP.Jphi);
	Torus T(TG.fitTorus(J));
	math::Matrix<double> M(3,3);
	M(0,0)=T.freqs.Omegar; M(1,0)=T.freqs.Omegaz; 
	M(2,0)=T.freqs.Omegaphi;
	J.Jr -= dJ;
	T=TG.fitTorus(J);
	M(0,0)-=T.freqs.Omegar; M(1,0)-=T.freqs.Omegaz; 
	M(2,0)-=T.freqs.Omegaphi;
	M(0,0)/=dJ; M(1,0)/=dJ; M(2,0)/=dJ;
	J.Jr += .5*dJ;
}
\noindent and similarly for the other rows. $\vM$ ought to be
symmetric but numerical errors introduce a small anti-symmetric component
which we eliminate by taking $M_{ij}$ to be the average of the originally
obtained values for $M_{ij}$ and $M_{ji}$. The eigenvectors and eigenvalues
of the symmetrised matrix are extracted by singular-value decomposition
  {\obeylines\tt\parindent=10pt 
	math::SVDecomp SVD(M);
	std::vector<double> lambdas(SVD.S()); 
	math::Matrix<double> U(SVD.U());
}

Let $\theta_\parallel=\ve_1\cdot\vtheta$ be the component of $\vtheta$
parallel to the eigenvector $\ve_1$ of $\vM$ with the largest eigenvalue. Then
stream extension reflects the steady growth in the differences
$\Delta\theta_\parallel$ between the $\theta_\parallel$ values of stars and
the progenitor.  Fig.~\ref{fig:simpleStream3} is a plot of drift time
$t_{\rm drft}$ against  the absolute
values of these differences. Fig.~5 of
\cite{Bovy2014:streams} shows the corresponding plot from an N-body
simulation. The two plots are similar: the main differences being (i) that
the horizontal streaks of particles released at a given pericentre shrink
towards the bottom left of the plot less fast in the model than in the N-body
data, and (ii) that clouds of particles are visible at the left end of the
streaks in the N-body data but not in the model. In the data streaks shrink
in length from top to bottom because the length of a streak is set by the
cluster's velocity dispersion and this decreases as the cluster is ablated.
In the model the dispersion is held constant, but it could be decreased by
gradually diminishing the parameter {\tt sigCluster} above. It's likely that
the particles that form clouds at the left ends of streaks are stars
that have not been liberated from the cluster but are on chaotic orbits
through the region where neither the cluster's field nor that of the
galaxy dominates.

\begin{figure}
\centerline{\includegraphics[width=\hsize]{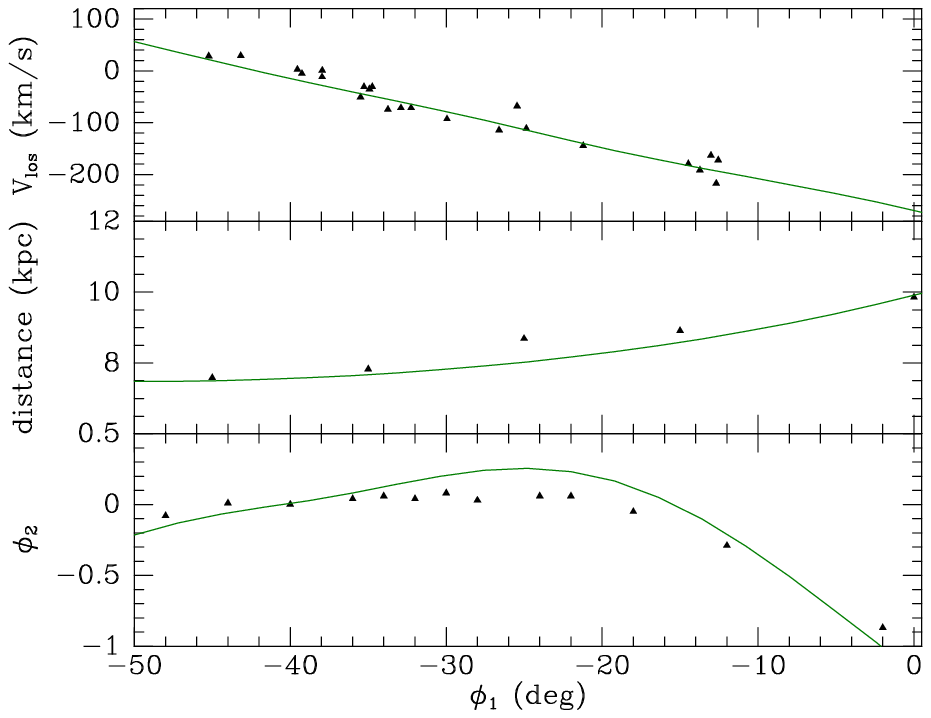}}
\caption{Data for the GD1 stream: from bottom to top, track on the sky,
distance as a function of angle $\phi_1$ down the stream, and line-of-sight
velocity. The black
triangles show data from Koposov et al (2010) while the green curves show the
cubic spline estimates of Valluri et al.~(2025). Since  $\mu_{\phi_1}<0$, stars move from right to
left and the leading part of the stream is on the left.}\label{fig:stream0}
\end{figure}

\begin{figure}
\centerline{\includegraphics[width=.8\hsize]{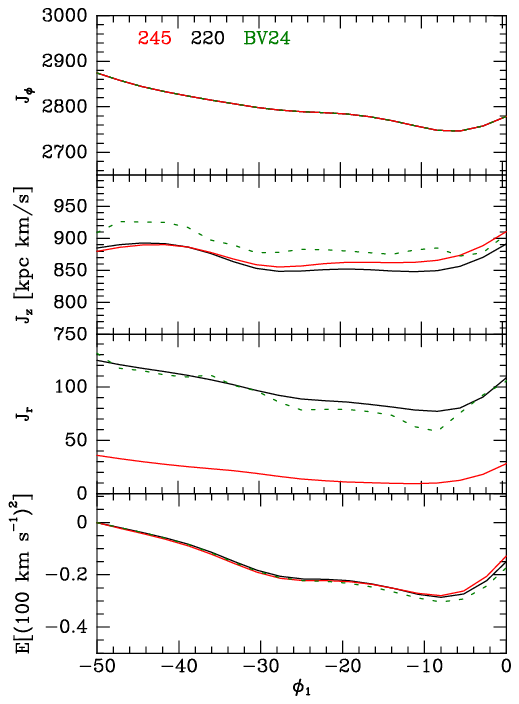}}
 \caption{Values of the actions and energy along the GD1 stream computed from
the cubic spline fits to the observations by Valluri et al.\ (2025) in three
potentials. The full black and red curves are for a flat circular speed curve
with $V_c=220\kms$ and $V_c=245\kms$, respectively.  The broken green curves
are for the potential inferred by Binney \& Vasiliev (2024). The zero point
in 
energy is set to the energy of the stream at $\phi_1=-50\,$ deg.}
\label{fig:plotGD1}
\end{figure}

\subsection{Application to GD1}

\cite{Valluri2025} give remarkably complete data for the long, thin GD1
stream, which was discovered by \cite{GrillmairDionatos2006}.  The green
curves in Fig.~\ref{fig:stream0} show the cubic spline fits to data from Gaia
DR3 and the Dark Energy Survey in \cite{Valluri2025}. The path across the sky
is given in terms of the coordinate system defined by \cite{KoRiHo09} in
which the stream keeps close to the equator. The curves given by
\cite{Valluri2025} are converted to conventional equatorial coordinates using
the transformation given by \cite{KoRiHo09} and then to conventional Galactic
coordinates $(\ell,b)$ by \agamaTwo's function {\tt obs::from\_RAdec}. The
method {\tt reflex} of {\tt solarShifter} is used to restore the
contributions to proper motions from the Sun's velocity, and then {\tt
solarShifter}'s method {\tt toCyl} converts the positions, proper motions,
distances and line-of-sight velocities to cylindrical phase-space
coordinates.  Finally {\tt totalEnergy} and {\tt actionFinderTG} are used to
compute energies and actions along the stream.

Fig.~\ref{fig:plotGD1} shows the resulting actions as functions of the angle
$\phi_1$ that runs along the stream computed in three gravitational
potentials. Two of these potentials have flat circular-speed curves (in the
sense of eqn \ref{eq:flatVc}), one with
$V_c=220\kms$ (black curves) and one with $V_c=245\kms$ (red curves). The
third potential is that obtained by \cite{BinneyVasiliev2024} by fitting Gaia
DR3 kinematics -- this potential makes $V_c$ a declining function of $R$ in
the relevant radius range.  Since $J_\phi$ is completely specified by the
data independently of the potential, only one curve is visible in the top
panel of Fig.~\ref{fig:plotGD1}. The next panel down shows weak dependence of $J_z$
on $\phi_1$, while the third  panel down shows significant
dependence of $J_r$ on $\phi_1$ -- increasing $V_c$ from $220\kms$ to $245\kms$
pushes $J_r$ down to remarkably small values -- for comparison, with
$V_c=220\kms$ the Sun's radial action would be $33\kpc\kms$. It seems that
the stream defines a nearly circular but highly inclined orbit.

The actions and energy should not vary much along a tidal stream.  From
$\phi_1=-50\,$deg to $-5\,$deg in Fig.~\ref{fig:plotGD1}  $J_\phi$ falls by
$\sim100\kpc\kms$ from $\sim2850\kpc\kms$ while $E$ drops by $\sim(30\kms)^2$,
which is equivalent to a star's speed changing by $\delta v\simeq\delta
E/v\sim0.1\kms$. Thus the constants of motion are indeed little changed along
the stream.  However, to the extent that the Galaxy's potential can be assumed
time-independent,  the sense of their changes is inconsistent with the values of
$\dot\phi_1$ along the stream: these values are negative, both in the raw
data and after subtraction of the contribution of the Sun's motion.  Hence
stars move along the stream from $\phi_1=0$ to $\phi_1=-50\,$deg and the
portion of leading arm of the stream is on the left in
Fig.~\ref{fig:plotGD1}.  The energy must be lowest,
and thus frequencies largest, in the leading arm, whereas the bottom panel of
Fig.~\ref{fig:plotGD1} reports the highest energies there.

\begin{figure}
\centerline{\includegraphics[width=.8\hsize]{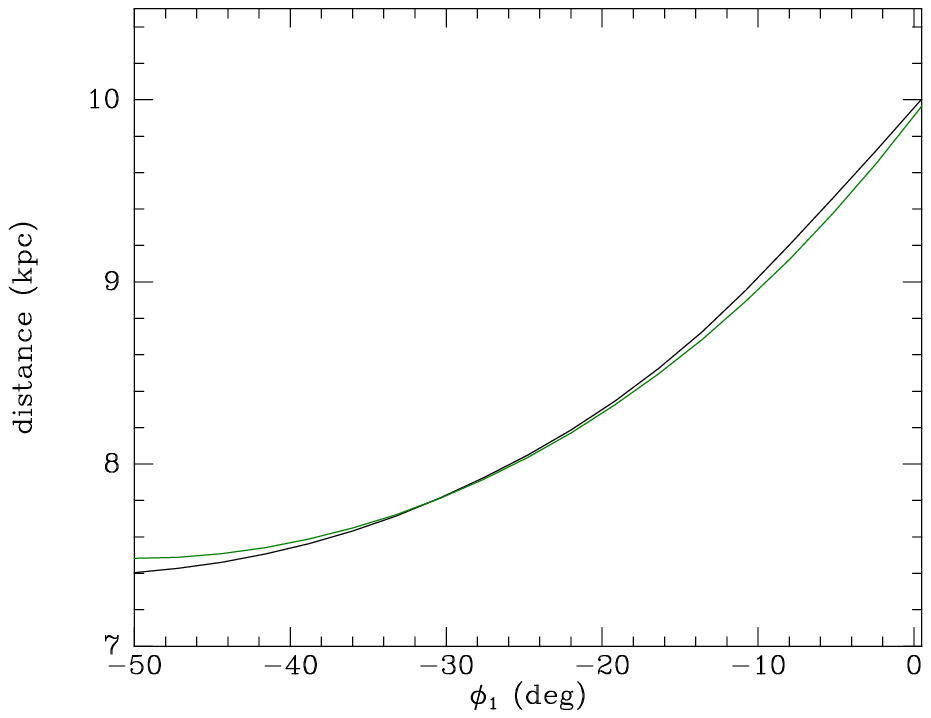}}
\caption{The sensitivity of recovered angular momenta and energies along GD1
to adopted distances. The green curve shows the distances given by Valluri
et al.\ (2025) while the black curve shows the distances that yield a
constant value $J_\phi=2800\kpc\kms$ along the stream.}\label{fig:constJphi}
\end{figure}

Our energies depend on assumed potentials, so does  this contradiction reflect
badly chosen potentials? The slope of the plot of $J_\phi$ at the top of
Fig.~\ref{fig:plotGD1} strongly suggests otherwise since higher frequencies
are associated with smaller $J_\phi$ as well as lower $E$ and $J_\phi$ does
not depend on the potential. We conclude that the problem 
lies either with the distances $s(\phi_1)$ we have adopted from
\cite{Valluri2025} or the stream being distorted by the tidal field of the
Magellanic Clouds \citep[e.g.][]{LMCtides2023}.

Fig.~\ref{fig:constJphi} suggests that the problem may lie with the adopted
distances by plotting in black the
distances $s(\phi_1)$ that yield a constant value $J_\phi=2800\kpc\kms$ along
the stream. We see that these distances differ insignificantly from those
given by \cite{Valluri2025}. With these distances the predicted variation in
energy along the stream is also much diminished: $E(\phi)$ oscillates with
amplitude $\la(20\kms)^2$ with no systematic drift along the
stream. 

\section{PYTHON wrappers}\label{sec:Python}

To exploit fully the possibilities offered by \agamaTwo, one should call its
routines from C++ code as illustrated above. However, we provide Python
wrappers of the underlying C++ code that allow many applications of
\agamaTwo\ to be completed from within Python. The wrappers take advantage of
the {\sc pybind11} package and differ substantially from those provided by
\cite{agama}. They are designed to facilitate input/output via Python while
leaving the computations and principal data structures in the C++ domain.

Python structures are provided that set up gravitational potentials,
distribution functions (DFs) and galaxy models (i.e., unions of DFs and a
potential). Wrappers are provided for the principal data types (actions,
angles, frequencies, phase-space and sky coordinates). With these tools,
Python can be used to set up a galaxy model or explore orbits in a potential
and then to plot  the resulting  observables (density, velocity
distributions, etc.) both in six-dimensional phase space and on the sky. 

As an example, with \agamaTwo\ imported as {\tt AG} and matplotlib imported
as {\tt plt} the following Python code will plot the trajectory across the
sky of an object like the progenitor of the GD1 stream:

  {\obeylines\tt\parindent=10pt 
IU=AG.galactic\_kms
sun=AG.solarShifter(IU)
h=IU.from\_Kpc\_kms
p=AG.createPotential("type=spheroid, gamma=1, beta=3, alpha=1, scaleradius=18, densityNorm=170, q=0.5")
TG=AG.TorusGenerator(p,1e-4)
J=AG.Actions(50*h,800*h,2800*h)
theta0=AG.Angles(0,0,4)
T=TG.fitTorus(J)
traj=T.orbit(theta0,1*IU.from\_Myr,.2*IU.from\_Gyr)
s=0
Vlos=0
l=[]
b=[]
for i in range(len(traj)):
\quad    astrom=sun.toSky(traj[i][0],s,Vlos)
\quad    l.append(astrom.pos.l)
\quad    b.append(astrom.pos.b)
plt.xlabel("longitude")
plt.ylabel("b")
plt.plot(l,b)
plt.show()
}

In this example {\tt IU} is an object that converts between internal and
astrophysical units: {\tt h} is then the unit of action and {\tt p} becomes
an NFW potential with scale radius $18\kpc$ and a density normalisation that
makes the circular speed at $8\kpc$ $246\kms$. The array {\tt traj} fills up
with {\tt PosVelCyl} in the slots {\tt traj[i][0]} and the corresponding
times in {\tt traj[i][1]}. {\tt sun.toSky} returns the corresponding
astrometry (and also computes distances and line-of-sight velocities, which
we discard).

\section{Conclusions}\label{sec:conclude}

We have presented a fork, \agamaTwo, of the {\sc agama} software suite
for action-based galaxy modelling. This includes a new torus mapper that uses
the data types and numerical procedures that are native to {\sc agama}, and
handles a wider range of orbital tori with greater economy and precision than
the BM16 torus code, wrappers for which were included in the original release
of {\sc agama} \citep{agama}. Significant upgrades include cleaner, more
effective point transformations and the use of angle-action coordinates from
both harmonic-oscillators and isochrones. The improved toy maps enable the new code to
construct tori that cannot be constructed by the old code. Construction of
tori by interpolation now works faultlessly. 

The new code includes an action finder that is more
accurate than the St\"ackel Fudge as well as a suite of functions for connecting
models to observational data.  

We have shown how orbits can be generated and used as either time series or
density distributions with a handful of statements. We have similarly
illustrated the generation of surfaces of section, both of the familiar
$(R,v_R)$ type and ones defined in toy angle-action space which can help
understand the structure of phase space. 

As an illustration of the practical use of tori, we showed how tidal streams
can be economically modelled with tori. The plethora of tidal streams that
have recently been catalogued in the Galactic halo \citep{Ibata2024} offer a
unique opportunity to constrain the Galaxy's potential and dark-matter
distribution, but to take advantage of this opportunity we must be able to
model observed streams in trial Galaxy potentials cheaply.
Section~\ref{sec:streams} showed that tori make it possible to reproduce in a
matter of seconds on a laptop a tidal stream produced by an N-body
integration. 

In a second exercise the new action finder was used to compute
actions along the GD1 stream using recently published observational data and
three trial potentials. The recovered actions vary slowly and within narrow
ranges as is to be expected for a stream, but energy and $J_\phi$ (which is
independent of the potential) appears to increase rather than decrease in the
direction of stellar motion, which is contrary to expectations. However,
negligible adjustments to the published distances suffice to make $J_\phi$
exactly constant along the stream and $E$ nearly constant. 

Section \ref{sec:streams} on streams illustrated the use of new \agamaTwo\
classes that facilitate comparisons between models and observational data.
These classes are more fully explained in Appendix~\ref{sec:obs} -- they
include facilities for switching between heliocentric and galactocentric
coordinates, sampling lines of sight both within our Galaxy and through
external galaxies, even in the presence of dust.

Much of the code can be accessed via Python through wrappers that are
illustrated in Section~\ref{sec:Python}.  Appendix~\ref{sec:behindScenes}
explains how tori are generated, both by minimising the variance of the
Hamiltonian and by interpolation, and explains the algorithm behind the new
action finder.

Classically, galaxy models have been constructed by two distinct routes: (i)
predicate a DF such as $f(E)$ or $f(E,L_z)$ and then find the corresponding
density $\rho(\vx)$ and kinematics by integrating over $\vv$
\citep[e.g.][]{King1966,Rowley1988}, and (ii) predicate the potential
$\Phi(\vx)$, integrate orbits in it and seek weights for the orbits that are
consistent with $\Phi$ and data \citep[e.g.][and references
therein]{Schwarzschild1979,MAGPI2024}. The first approach has been extended
to replace $E$ by actions \citep[e.g.][]{JJB14,BinneyPiffl15,BinneyVasiliev2024}
and the time is ripe to adapt the second approach to replace orbits by tori.
Indeed, the {\tt density} and {\tt containsPoint} methods of the class {\tt
Torus} makes Schwarzschild modelling a natural application of tori. It is
easier to sample orbit space systematically by constructing tori at the nodes
of a three-dimensional grid than by integrating orbits from the nodes of a
grid of points in six-dimensional phase space, and tori are much more
economically
stored than the long integrations that are necessary to ensure adequate
coverage of phase space. Moreover, a torus-based model's resolution can be
enhanced by interpolating between the grid's tori, and the final model is
easier to interpret if it comprises the weights of tori on a regular grid in
action space than the weights of orbits. Notwithstanding these apparent
advantages, to date all action-based modelling has been based on specified,
analytic DFs rather than on orbit/torus weights.  The opportunities and
pitfalls associated with models that are described by a large number of orbit
weights are quite different from those associated with analytic DFs, so
Schwarzschild modelling with tori should be tried.

Torus mapping yields an integrable Hamiltonian that is  as close as any to a
given Hamiltonian, and thus provides a unique framework for perturbation
theory \citep{Kaasalainen_res}. \cite{Binney2016} introduced the {\it rTorus}
class to the old torus-mapping code library \citep{JJBPJM16} and
\cite{Binney2018} used this
class to study orbit trapping by the Galactic bar. When \agamaTwo\ has been
extended to include a class  {\it rTorus}, it promises to be a powerful probe of the
role of resonant trapping in galactic dynamics.

\section*{Acknowledgements}
 We thank Hongyi Xiong for insightful comments, Monica Valluri for clarifying
the contents of a paper and Rodrigo Ibata for the provision of Gaia data for
GD1. The code that computes angle variables and frequency was originally
written in 2015 by Brian Khor while an undergraduate summer intern.  TW was
supported by the Leverhulme Trust under grant LIP-2020-014, while EV
acknowledges support from an STFC Ernest Rutherford Fellowship
(ST/X0040066/1). JB and EV were in part supported by grant no.  NSF PHY-2309135
to the Kavli Institute for Theoretical Physics (KITP).

\section*{DATA AVAILABILITY}

The code presented here and
available at {\tt https://github.com/binneyox/AGAMAb} must be
considered a distinct branch, \agamaTwo, of {\sc agama} from that available at 
{\tt https://github.com/GalacticDynamics-Oxford/Agama}. Ideally the two
code branches would be merged, but a substantial effort would be required and
we have no immediate plans to undertake this.
Whatever your operating system, clone the repository. On a Linux machine then
just move to the {\tt src} directory and run `pip install .'. Much of code
development  has been done on a Windows machine using the free MS Visual
Studio compiler under the associated IDE. A document {\it agama\_vs.pdf} in
the repository explains what's involved in getting \agamaTwo\ to work in the
Visal Studio IDE.

\def\physrep{Phys.~Reps}
\def\apss{Ast.~Phys.~Sp.~Sci.}
\def\fcp{Fund.~Cosmic~Phys.}
\bibliographystyle{mn2e} \bibliography{/u/tex/papers/mcmillan/torus/new_refs}

\begin{thebibliography}{50}
\expandafter\ifx\csname natexlab\endcsname\relax\def\natexlab#1{#1}\fi

\bibitem[{{Binney}(2008)}]{JJBstream2008}
{Binney} J., 2008, \mnras, 386, L47

\bibitem[{{Binney}(2012)}]{JJB12:Stackel}
{Binney} J., 2012, \mnras, 426, 1324

\bibitem[{{Binney}(2014)}]{JJB14}
{Binney} J., 2014, \mnras, 440, 787

\bibitem[{{Binney}(2016)}]{Binney2016}
{Binney} J., 2016, \mnras, 462, 2792

\bibitem[{{Binney}(2018)}]{Binney2018}
{Binney} J., 2018, \mnras, 474, 2706

\bibitem[{{Binney}(2020)}]{Binney_negJ}
{Binney} J., 2020, \mnras, 495, 886

\bibitem[{{Binney}(2024)}]{BinneyWarp2024}
{Binney} J., 2024, \mnras, 535, 1898

\bibitem[{{Binney} \& {Kumar}(1993)}]{JJBKu93}
{Binney} J., {Kumar} S., 1993, \mnras, 261, 584

\bibitem[{{Binney} \& {McMillan}(2016)}]{JJBPJM16}
{Binney} J., {McMillan} P.~J., 2016, \mnras, 456, 1982

\bibitem[{{Binney} \& {Piffl}(2015)}]{BinneyPiffl15}
{Binney} J., {Piffl} T., 2015, \mnras, 454, 3653

\bibitem[{{Binney} \& {Tremaine}(2008)}]{GDII}
{Binney} J., {Tremaine} S., 2008, {Galactic Dynamics: Second Edition}.
  Princeton University Press

\bibitem[{{Binney} \& {Vasiliev}(2023)}]{BinneyVasiliev2023}
{Binney} J., {Vasiliev} E., 2023, \mnras, 520, 1832

\bibitem[{{Binney} \& {Vasiliev}(2024)}]{BinneyVasiliev2024}
{Binney} J., {Vasiliev} E., 2024, \mnras, 527, 1915

\bibitem[{{Bovy}(2014)}]{Bovy2014:streams}
{Bovy} J., 2014, \apj, 795, 95

\bibitem[{{Cappellari}(2016)}]{Cappellari2016}
{Cappellari} M., 2016, \araa, 54, 597

\bibitem[{{Cappellari}(2011)}]{ATLAS3D_short}
{Cappellari} M. e.~a., 2011, \mnras, 413, 813

\bibitem[{{de Zeeuw}(1985)}]{deZeeuw1985}
{de Zeeuw} T., 1985, \mnras, 216, 273

\bibitem[{{Erkal} {et~al}\mbox{.}(2019){Erkal}, {Belokurov}, {Laporte},
  {Koposov}, {Li}, {Grillmair}, {Kallivayalil}, {Price-Whelan}, {Evans},
  {Hawkins}, {Hendel}, {Mateu}, {Navarro}, {del Pino}, {Slater}, {Sohn}, \&
  {Orphan Aspen Treasury Collaboration}}]{Erkal2019}
{Erkal} D. {et~al.}, 2019, \mnras, 487, 2685

\bibitem[{{Eyre} \& {Binney}(2011)}]{EyJJB11}
{Eyre} A., {Binney} J., 2011, \mnras, 413, 1852

\bibitem[{{GRAVITY Collaboration} {et~al}\mbox{.}(2022){GRAVITY Collaboration},
  {Abuter}, {Aimar}, {Amorim}, {Ball}, {Baub{\"o}ck}, {Berger}, {Bonnet},
  {Bourdarot}, {Brandner}, {Cardoso}, {Cl{\'e}net}, {Dallilar}, {Davies}, {de
  Zeeuw}, {Dexter}, {Drescher}, {Eisenhauer}, {F{\"o}rster Schreiber},
  {Foschi}, {Garcia}, {Gao}, {Gendron}, {Genzel}, {Gillessen}, {Habibi},
  {Haubois}, {Hei{\ss}el}, {Henning}, {Hippler}, {Horrobin}, {Jochum}, {Jocou},
  {Kaufer}, {Kervella}, {Lacour}, {Lapeyr{\`e}re}, {Le Bouquin}, {L{\'e}na},
  {Lutz}, {Ott}, {Paumard}, {Perraut}, {Perrin}, {Pfuhl}, {Rabien},
  {Shangguan}, {Shimizu}, {Scheithauer}, {Stadler}, {Stephens}, {Straub},
  {Straubmeier}, {Sturm}, {Tacconi}, {Tristram}, {Vincent}, {von Fellenberg},
  {Widmann}, {Wieprecht}, {Wiezorrek}, {Woillez}, {Yazici}, \&
  {Young}}]{Gravity2022}
{GRAVITY Collaboration} {et~al.}, 2022, \aap, 657, L12

\bibitem[{{Grillmair} \& {Dionatos}(2006)}]{GrillmairDionatos2006}
{Grillmair} C.~J., {Dionatos} O., 2006, \apjl, 643, L17

\bibitem[{{Helmi} \& {White}(1999)}]{HeWh99}
{Helmi} A., {White} S.~D.~M., 1999, \mnras, 307, 495

\bibitem[{{H\'enon}(1959)}]{Henon1959}
{H\'enon} M., 1959, Annales d'Astrophysique, 22, 126

\bibitem[{{H{\'e}non}(1960)}]{Henon1960}
{H{\'e}non} M., 1960, Annales d'Astrophysique, 23, 474

\bibitem[{{Hernquist}(1990)}]{He90}
{Hernquist} L., 1990, ApJ, 356, 359

\bibitem[{{Ibata} {et~al}\mbox{.}(2001){Ibata}, {Lewis}, {Irwin}, {Totten}, \&
  {Quinn}}]{Ibea01}
{Ibata} R., {Lewis} G.~F., {Irwin} M., {Totten} E., {Quinn} T., 2001, \apj,
  551, 294

\bibitem[{{Ibata} {et~al}\mbox{.}(2024){Ibata}, {Malhan}, {Tenachi},
  {Ardern-Arentsen}, {Bellazzini}, {Bianchini}, {Bonifacio}, {Caffau},
  {Diakogiannis}, {Errani}, {Famaey}, {Ferrone}, {Martin}, {di Matteo},
  {Monari}, {Renaud}, {Starkenburg}, {Thomas}, {Viswanathan}, \&
  {Yuan}}]{Ibata2024}
{Ibata} R. {et~al.}, 2024, \apj, 967, 89

\bibitem[{{Jin} {et~al}\mbox{.}(2024){Jin}, {Zhu}, {Zibetti}, {Costantin}, {van
  de Ven}, \& {Mao}}]{CALIFA2024}
{Jin} Y., {Zhu} L., {Zibetti} S., {Costantin} L., {van de Ven} G., {Mao} S.,
  2024, \aap, 681, A95

\bibitem[{{Julian} \& {Toomre}(1966)}]{JT1966}
{Julian} W.~H., {Toomre} A., 1966, \apj, 146, 810

\bibitem[{{Kaasalainen}(1994)}]{Kaasalainen_res}
{Kaasalainen} M., 1994, \mnras, 268, 1041

\bibitem[{{Kaasalainen}(1995)}]{Ka95:closed}
{Kaasalainen} M., 1995, \mnras, 275, 162

\bibitem[{{Kaasalainen} \& {Binney}(1994)}]{KaJJB94:MNRAS}
{Kaasalainen} M., {Binney} J., 1994, \mnras, 268, 1033

\bibitem[{{King}(1966)}]{King1966}
{King} I.~R., 1966, \aj, 71, 64

\bibitem[{{Koposov} {et~al}\mbox{.}(2010){Koposov}, {Rix}, \&
  {Hogg}}]{KoRiHo09}
{Koposov} S.~E., {Rix} H.-W., {Hogg} D.~W., 2010, \apj, 712, 260

\bibitem[{{Laakso} \& {Kaasalainen}(2013)}]{LaaksoKaas}
{Laakso} T., {Kaasalainen} M., 2013, Physica D Nonlinear Phenomena, 243, 14

\bibitem[{{Lilleengen} {et~al}\mbox{.}(2023){Lilleengen}, {Petersen}, {Erkal},
  {Pe{\~n}arrubia}, {Koposov}, {Li}, {Cullinane}, {Ji}, {Kuehn}, {Lewis},
  {Mackey}, {Pace}, {Shipp}, {Zucker}, {Bland-Hawthorn}, {Hilmi}, \& {S5
  Collaboration}}]{LMCtides2023}
{Lilleengen} S. {et~al.}, 2023, \mnras, 518, 774

\bibitem[{{McGill} \& {Binney}(1990)}]{McGJJB90}
{McGill} C., {Binney} J., 1990, \mnras, 244, 634

\bibitem[{{Press} {et~al}\mbox{.}(1992){Press}, {Flannery}, \&
  {Teukolsky}}]{NumRec}
{Press} W.~H., {Flannery} B.~P., {Teukolsky} S.~A., 1992, {Numerical recipes in
  C: the art of scientific computing}. Cambridge: University Press

\bibitem[{{Reid} \& {Brunthaler}(2020)}]{ReidBrunthaler2020}
{Reid} M.~J., {Brunthaler} A., 2020, \apj, 892, 39

\bibitem[{{Rowley}(1988)}]{Rowley1988}
{Rowley} G., 1988, \apj, 331, 124

\bibitem[{{Sanders}(2014)}]{Sa14}
{Sanders} J.~L., 2014, \mnras, 443, 423

\bibitem[{{Sanders} \& {Binney}(2013)}]{SaJJB13b}
{Sanders} J.~L., {Binney} J., 2013, \mnras, 433, 1826

\bibitem[{{Santucci} {et~al}\mbox{.}(2024){Santucci}, {Lagos}, {Harborne},
  {Derkenne}, {Poci}, {Thater}, {McDermid}, {Mendel}, {Wisnioski}, {Croom},
  {Ferr{\'e}-Mateu}, {Muller}, {van de Sande}, {Sharma}, {Sweet}, {Tsukui},
  {Valenzuela}, {van de Ven}, \& {Zafar}}]{MAGPI2024}
{Santucci} G. {et~al.}, 2024, \mnras, 534, 502

\bibitem[{{Sch{\"o}nrich}(2012)}]{Schoenrich2012}
{Sch{\"o}nrich} R., 2012, \mnras, 427, 274

\bibitem[{{Schwarzschild}(1979)}]{Schwarzschild1979}
{Schwarzschild} M., 1979, \apj, 232, 236

\bibitem[{{Tremaine}(1999)}]{Tr99}
{Tremaine} S., 1999, \mnras, 307, 877

\bibitem[{{Valluri} {et~al}\mbox{.}(2025){Valluri}, {Fagrelius}, {Koposov},
  {Li}, {Gnedin}, {Bell}, {Carlberg}, {Cooper}, {Aguilar}, {Ahlen}, {Allende
  Prieto}, {Belokurov}, {Beraldo e Silva}, {Brooks}, {Bystr{\"o}m},
  {Claybaugh}, {Dawson}, {Dey}, {Doel}, {Forero-Romero}, {Gazta{\~n}aga},
  {Gontcho A Gontcho}, {Han}, {Honscheid}, {Kisner}, {Kremin}, {Lambert},
  {Landriau}, {Le Guillou}, {Levi}, {de la Macorra}, {Manera}, {Martini},
  {Medina}, {Meisner}, {Miquel}, {Moustakas}, {Myers}, {Najita}, {Poppett},
  {Prada}, {Rezaie}, {Rossi}, {Riley}, {Sanchez}, {Schlegel}, {Schubnell},
  {Sprayberry}, {Tarl{\'e}}, {Thomas}, {Weaver}, {Wechsler}, {Zhou}, \&
  {Zou}}]{Valluri2025}
{Valluri} M. {et~al.}, 2025, \apj, 980, 71

\bibitem[{{Vasiliev}(2019)}]{agama}
{Vasiliev} E., 2019, \mnras, 482, 1525

\bibitem[{{Wright} \& {Binney}(2025)}]{WrightB}
{Wright} T., {Binney} J., 2025, arXiv, 2512.06519

\bibitem[{{Zhu} {et~al}\mbox{.}(2018){Zhu}, {van de Ven}, {M{\'e}ndez-Abreu},
  \& {Obreja}}]{Zhu2018}
{Zhu} L., {van de Ven} G., {M{\'e}ndez-Abreu} J., {Obreja} A., 2018, \mnras,
  479, 945

\end{thebibliography}

\appendix

\section{What goes on behind the scenes}\label{sec:behindScenes}

This section explains how \agamaTwo\ creates tori and uses them in the new
action finder before closing with some plots that are useful in understanding
how well the code is working. Readers who are only interested in
producing and using tori can skip this section.

When an {\it actionFinder} is initialised in \agama, shell orbits are
integrated on a grid in energy and $\xi\equiv J_\phi/L_{\rm circ}$ and values
are stored of each orbit's intercept $\Rsh$ with the equatorial plane, and
its focal distance $\Delta$. Given that these data are also required by a
{\it TorusGenerator}, in \agamaTwo\ they are associated with a potential
rather than an {\it actionFinder} and they are normally computed when a
potential is initialised. Such a potential has methods that return $\Rsh$ and
$\Delta$ as functions of either $(E,\xi)$ or $(L,\xi)$: the former dependence
is required by an {\it actionFinder} while the latter is what's needed by a
{\it TorusGenerator}. 

At each energy $E$, $J_\phi$ is decreased from its circular value. During
the orbit integration $J_z$ is determined by integrating the differential
equation $\d J/\d t=\vp\cdot\vv$ in parallel with the equations of motion.
Then to each orbit we can attach the numbers
\[
L\equiv J_z+|J_\phi| \quad\hbox{and} \quad \xi\equiv {|J_\phi|\over L}
\]
-- in the case of a spherical potential, $\xi=\cos i$ is a measure of the
orbit's inclination $i$.  Each shell orbit is fitted with an ellipse
with minor semi-axis length $\Rsh$ and focal distance $\Delta$. From
these data interpolators are constructed from which $\Rsh$ and $\Delta$
can be predicted at either arbitrary $(L,\xi)$ or $(E,|J_\phi|/L_{\rm
circ})$.  The interpolation grid typically
has $\sim100$ points in $L$ and 20 in $\xi$ with non-uniform spacing in each
direction.

Once the survey of shell orbits is complete, \agamaTwo\ computes box/loop
transition orbits on a grid in energy and stores their values of $J_z$ as
both functions of $E$ and $J_{\rm fast}\equiv 2J_r+J_z$, At the same time
values of three quantities, $I_{3\rm crit}$, $\Delta_{\rm crit}$ and $u_{\rm
min}$, are stored for these orbits as functions of $E$. These data are used
to refine the St\"ackel Fudge as explained in \cite{WrightB}.

When a {\it TorusGenerator} is initialised, one more table is constructed.
This gives the energy of orbits in the equatorial plane  as a function of 
$J_r+|J_\phi|$ and $|J_\phi|/(J_r+|J_\phi|)$. The table is stored as a quintic
spline {\tt interpJrE}.

\subsection{Choosing an HJ map}\label{sec:ToyMap}

The first task after a call of {\tt fitTorus} is the choice of an HJ map. It
is undertaken by {\it TorusGenerator}'s method {\tt chooseTM}. It starts by
using $L$ and $\xi$ to obtain by
interpolation the radius $\Rsh$ and focal distance $\Delta$ of the
underlying shell orbit. $\Rsh$ sets the characteristic frequency
$\Omega_{\rm scale}$ in the convergence criterion
\[\label{eq:converge}
\sigma_H<{\tt tol}\times\Omega_{\rm scale}J_{\rm scale}.
\]

Next the table {\tt interpJrE} is used to estimate the
energy from $\vJ$. This is used to determine the orbit's velocity $v_{\rm
merid}$ in the
meridional plane when the star is at $(R,z)=(\Delta,0)$, and the variable
$J_{\rm low}=0.3\Delta v_{\rm merid}$ is set. If $J_\phi>J_{\rm low}$, an
isochrone HJ map is selected. If $J_\phi<J_{\rm low}$, $\Jcrit$ is
evaluated from the potential's interpolator and a harmonic-oscillator HJ map
is selected if $J_z<\Jcrit$ and an isochrone otherwise.

\subsubsection{Choosing an isochrone}

An isochrone is defined by the numbers $\Js$ and $b$. 
The condition for the isochrone's circular orbit with $L=J_z+|J_\phi|$ to have
radius $\Rsh$ can be written
\[\label{eq:RtoJs}
\Rsh=bg(\Js),
\]
where
\[
g(\Js)\equiv\sqrt{{\Js^4\over(2b^2\cE_c)^2}-1}.
\]
Here $-\cE_c$ is the energy of the circular orbit, which satisfies
\[
b^2\cE_c={2\Js^4\over(L+\sqrt{L^2+4\Js^2})^2}.
\]
 Equation (\ref{eq:RtoJs}) defines a one-parameter family of possible
isochrones. We choose among this family by requiring that the ratio of the
radial forces at peri- and apo-centre match the corresponding ratio in the
real potential. Specifically, for a trial value $\Js$ and the corresponding
$b=\Rsh/g(\Js)$ we evaluate
\[
b^2\cE\equiv{2\Js^4\over[2J_r+L+\sqrt{L^2+4\Js^2}]^2}
\]
which is $b^2$ times the absolute value of the energy of the isochrone orbit,
and then the dimensionless parameter
\[
\gamma\equiv{\Js^2\over2b^2\cE}-1.
\]
The orbit's eccentricity is
\[
e(\Js)=\sqrt{1-{L^2\over\gamma\Js^2}\Big(1+{1\over\gamma}\Big)}
\]
and from $e$ the peri- and apo-radii follow
\[
r_\pm(b,\Js)=\gamma b\sqrt{(1\pm e)(1\pm e+2/\gamma)}.
\]
The condition on the radial forces can now be written
\[
{(b+a_-)^2a_-r_+\over(b+a_+)^2a_+r_-}={(\p\Phi/\p R)_{(r_+,0)}\over(\p\Phi/\p
R)_{(r_-,0)}}
\]
 where $a(r)=\sqrt{b^2+r^2}$ and $a_\pm\equiv a(r_\pm)$. An instance of the
class {\it JsFinder} solves this equation for $\Js$ with $b$ an explicit
function of $\Js$ through equation (\ref{eq:RtoJs}). The isochrone is fully
specified once $\Js$ and $b$ have been chosen.

\subsubsection{Choosing a harmonic oscillator}

The radius of a degenerate two-dimensional harmonic oscillator of natural
frequency $\omega_R$ is given by
\[
R^2={E\over\omega_R^2}(1+\beta\cos\theta_r),
\]
where $E=\omega_R(2J_r+|J_\phi|)$ and $\beta=\sqrt{1-(\omega_R J_\phi/E)^2}$. So
to set its apo-centric radius to $R_{\rm max}$,  we need
frequency
\[\label{eq:2dHO}
\omega_R={1\over R_{\rm max}^2}(2J_r+|J_\phi|)(1+\beta).
\]
The amplitude $Z$ of the vertical oscillator is given by $Z^2=2J_z/\omega_z$.
In the case of a near-shell orbit it's also related to $\Rsh$ and the
inclination  $\sim J_z/L$. We adopt
\[
Z={2J_z\over\pi L}\Rsh.
\]
 Solving for $\omega_z$ with this value of $Z$ yields
\[
\omega_z={\pi^2\over2}{L^2\over\Rsh^2J_z}.
\]

\subsection{Point transformation}\label{sec:PtTrans}

The output $(r,\vartheta,p_r,p_\vartheta)$ or $(R',z',p_R',p_z')$ from the chosen
HJ map becomes the input to a point transformation
$(r,\vartheta)\leftrightarrow(R,z)$ or $(R',z')\leftrightarrow(R,z)$ to
determine the location $(R,z)$ at which the real potential will be evaluated.

Given a radial variable such as $x=r$ or $x=R'$ that lies in $(0,\infty)$, we define
a dimensionless variable that is  confined to $(\fracj12,1)$ by
 \[\label{eq:def_s}
s=\fracj12\big[1+\tanh(x/x_0)\big],
\]
 where $x_0$ is a suitable linear scale.  Then we transform $x$ to a new
dimensional variable
\[\label{eq:def_xn}
\widetilde x=x+x_0\sum_{n=1}^{N}X^{(x)}_n\sin(2n\pi s).
\]
 where the $X^{(x)}_n$ are dimensionless constants chosen as described below.
The derivatives $\d \widetilde x/\d x$ and $\d^2\widetilde x/\d x^2$ are
obtained analytically. Satisfactory results are obtained with $N=5$ terms in
the Fourier series.

When using an isoschrone as the HJ map, the map
$(r,\vartheta)\leftrightarrow(R,z)$ is as follows.  We write
\begin{align}\label{eq:PTtrans}
R=\widetilde r\sin v\quad;\quad z=\sqrt{\widetilde r^2+\Delta^2}\cos v,
\end{align}
 where $v(\vartheta)$ is defined by the Fourier series
\[\label{eq:vFourier}
v(\vartheta)=\vartheta+\sum_{n=1}^NV^{(\vartheta)}_n\sin(2n\vartheta).
\]
The inverse map is given by
\begin{align}\label{eq:inverse}
\widetilde r^2&=\fracj12\big(B+\sqrt{B^2+4R^2\Delta^2}\big)\cr
\cos v&={z\over\sqrt{\widetilde r^2+\Delta^2}},
\end{align}
 where $B=R^2+z^2-\Delta^2$. From $\widetilde r$ we obtain $r$ via equations
(\ref{eq:def_s}) and (\ref{eq:def_xn}) with $x\to r$. 

The point transformation (\ref{eq:PTtrans})
maps circles $\widetilde r=\hbox{const}$ into ellipses that have foci on the $z$ axis at
$z=\pm\Delta$. The function $v(\vartheta)$ determines how each circle is
mapped to its ellipse.  The corresponding inclined
circular orbit of the isochrone maps to the
ellipse defined by equation (\ref{eq:ellipse}) and thus nearly coincides
with the real-space trace of the galaxy's shell orbit because 
$\Delta$ was determined by fitting  an
ellipse to the shell orbit $\vJ=(0,J_z,J_\phi)$.  

When using a harmonic-oscillator HJ map, $R'$ is mapped to $R$ by setting
$x=R'$ and $\widetilde x=R$ in equations (\ref{eq:def_s}) and
(\ref{eq:def_xn}).  Equation (\ref{eq:def_s}) is used to map $z'$ to $s$ but
then $s$ is mapped to an angular variable $v$ rather than to an analogue of
$\widetilde x$. Specifically we define $v(s)$ by
\[
v=\pi(1-s)+\sum_{n=1}^NV_n^{(z)}\sin\big(2n\pi s\big).
\] 
 As $z'$ increases from zero to large values, $1-s$ decreases from $\pi/2$
towards zero like the angle $\vartheta$ of spherical polar coordinates.
Finally, the second of equations (\ref{eq:inverse}) with $\widetilde
r\to\widetilde R$ is used
to obtain $z$. 

In summary, an isochrone map is combined with a point transformation
$(r,\vartheta)\leftrightarrow(R,z)$, while a harmonic-oscillator map is
combined with a map $(R', z')\leftrightarrow(R,z)$. The role of the Fourier
coefficients in equations (\ref{eq:def_xn}) and (\ref{eq:vFourier}) is to
ensure that when one action vanishes and the orbit reduces to a curve in real
space and a loop in position-momentum space, the image of the toy orbit in
position-momentum space fits the trace  in this space of the real orbit.
Failure of the point transformation to provide a good fit is likely to cause  the
generating function $S(\vJ,\vthetaT)$ (eqn.~\ref{eq:defS}) to yield negative toy
actions, which are meaningless.

\begin{figure}
\centerline{\includegraphics[width=.8\hsize]{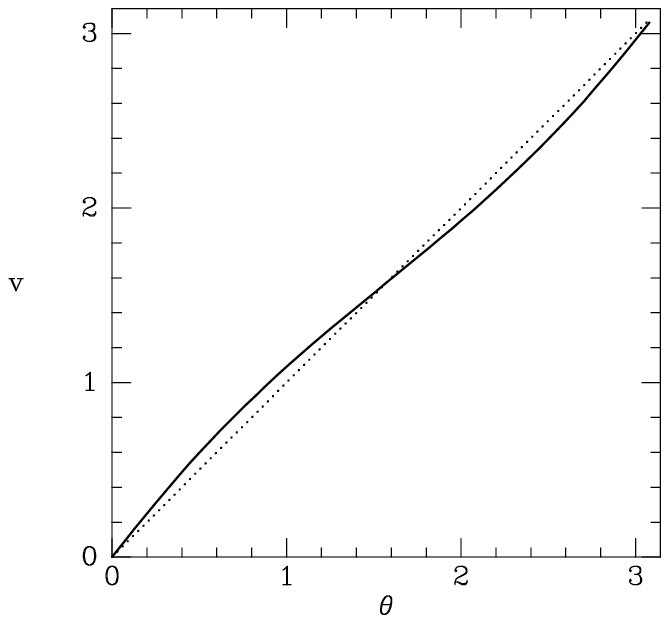}}
\caption{The full curve shows the relationship $v(\vartheta)$ between the
cylindrical and spherical angles that maps the $(\vartheta,p_\vartheta)$ trace of a circular orbit in the isochrone
potential onto the $(z,p_z)$ trace of a shell orbit in the given flattened
potential -- see text for details. The dotted line shows $v=\vartheta$.} \label{fig:thetaPsi}
\end{figure}

Fig.~\ref{fig:thetaPsi} is a plot of $v(\vartheta)$ for the orbit plotted in
Figs.~\ref{fig:t-seqs} to \ref{fig:Xsecs}. The relation is
close to the identity $v=\vartheta$ but the deviation from the identity
dramatically improves the performance of the {\it TorusGenerator} --
Fig.~\ref{fig:JrSoS} below illustrates this fact. 

Toy-map selection finishes with determination of the Fourier coefficients
$X_n$ and $V_n$ (eqns \ref{eq:def_xn} and \ref{eq:vFourier}). The adopted
values are those that minimise the variance in $H$ around the torus produced
by the toy map acting alone -- the Levenberg-Marquardt algorithm determines
the coefficients.

\subsection{Determining the $S_\vk$}\label{sec:Sk}

\begin{figure}
\centerline{\includegraphics[width=.8\hsize]{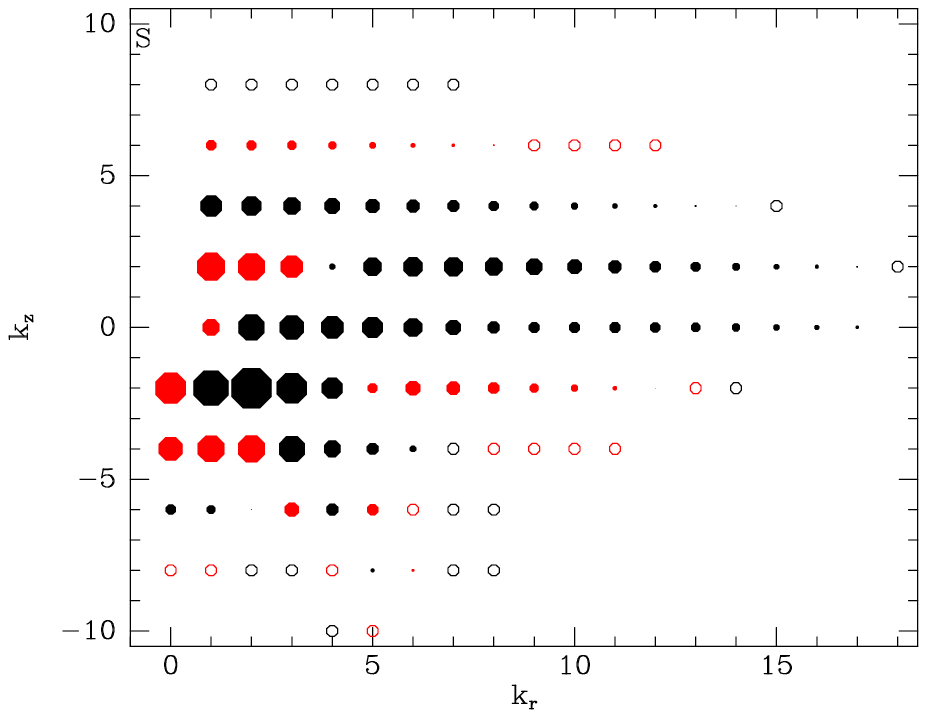}}
 \caption{The structure of the generating function of the orbit plotted in
Fig.~\ref{fig:t-seqs}.  The radius of
each dot is proportional to the logarithm of the corresponding $|S_\vk|$ in
the generating function (\ref{eq:defS}) and the point is black or red
according as $S_\vk$ is positive or negative. An open circle indicates that
the term was included in the optimisation but its value is too small to make
a visible dot.}\label{fig:Spts}
\end{figure}

Once the toy map has been defined, the code chooses the integer vectors $\vk$
that will label non-zero $S_\vk$ in the generating function (\ref{eq:defS}).
If the potential is even in $z$, only vectors with even $k_z$ are required
(BM16). Moreover, the symmetry of the generating function (\ref{eq:thT2th})
is such that if $\vk$ is included, $-\vk$ is not required. We implement this
condition by selecting only vectors with $k_r\ge0$ and omitting vectors with
$k_r=0$ and $k_z\ge0$.  Fig.~\ref{fig:Spts} illustrates this situation. By
default terms are included up to $k_r=6$ and $|k_z|=4$. The
Levenberg-Marquardt algorithm now chooses the $S_\vk$ to minimise the
variance of $H$.

If the dispersion in $H$ does not satisfy the convergence criterion
(\ref{eq:converge}), additional terms are added to the generating function.
The boundary of the region of non-zero $S_\vk$ is inspected, and new
coefficients are declared wherever a point adjacent to the boundary has a
neighbour whose magnitude exceeds 20 per cent of the magnitude of the largest
point on the current boundary. `Neighbour' is defined to be differences in
$k_r$ no more than three and $k_z$ no more than two. Once the list of $\vk$
vectors
has been expanded, the values of the $S_\vk$ are again optimised, starting
from the previous values in the case of coefficients that are not new.
Fig.~\ref{fig:Spts} shows the generating function of the orbit plotted in
Fig.~\ref{fig:t-seqs}. 

Once the convergence criterion has been satisfied (or an increase in the
number of $S_\vk$ has not reduced $\sigma_H$ significantly), equations
(\ref{eq:giveOmega}) are solved for the frequencies $\Omega_i$ and
derivatives $\p S_\vk/\p J_i$.  

It's important to sample tori sufficiently densely in $\vthetaT$, both when
minimising the variance in $H$ and when solving equations
(\ref{eq:giveOmega}). When a generating function is set up, the largest value
$k_{i\rm max}$ is determined for each $i$, and the nodes of the angle grid
are set at $\thetaT_i=j\pi/K_i$ for $j=0,\ldots,K_i-1$, where
$K_i\simeq1.5(k_{i\rm max}+1)$. There are of order $2k_{r\rm max}k_{z\rm
max}$ coefficients $S_\vk$ to optimise, so when minimising the variance in
$H$, we require $K_rK_z>2k_{r\rm max}k_{z\rm max}$. A slightly more stringent
criterion applies when solving for the $\p S_\vk/\p J_i$ because then we have
to add $\Omega_i$ to the list of unknowns to be determined from the $i$th
equation.

The final step of torus generation is to package the $S_\vk$ and their
derivative into an instance of the {\it GenFnc} class. In addition to holding
the $S_\vk$, this class provides methods for moving between true and toy
actions and angles, and for evaluating Jacobian matrices.  It also has {\it
read} and {\it write} methods so generating functions (and tori) can be
filed.

\subsection{Interpolation}

The basic building blocks of a torus are a generating function
and a toy map.
Functions such as {\tt
interpGenFnc}, and {\tt interpPtrToyMap} are defined that linearly
interpolate between two instances of either of these classes. For example 
  {\obeylines\tt\parindent=10pt 
GenFnc GF2 = interpGenFnc(x, GF0, GF1);
}
\noindent where $0\le x\le1$ makes {\tt GF2} $x$ times {\tt GF0} plus $1-x$ times
{\tt GF1}. 

\noindent 
Some terms $S_\vk$ may be in one generating function but not in the other. The
interpolated generating function contains all the terms that appear in either
of the input generating functions, with zero used as the amplitude of a
missing term. Perturbing Hamiltonians are interpolated in the same way.

Interpolation between toy maps is straightforward and can be accomplished by
a function {\tt InterpTorus(x,T0,T1)} so long as the tori use the same HJ
map. If one torus uses an isochrone and the other uses a harmonic oscillator,
interpolation must be achieved by the method {\tt interpTorus} of {\it
TorusGenerator}. This method replaces the torus that uses an isochrone by one
that uses a harmonic oscillator before calling {\tt InterpTorus}.
 
In Section~\ref{sec:streams} we used  the class {\tt TorusGrid3} that
provides interpolation on three-dimensional grids of tori.
Resonances drive orbits along one-dimensional manifolds in action space that all have
the same energy by virtue of the resonant condition 
\[
0={\vN\cdot\vOmega}=N_r{\p H\over\p J_r}+N_z{\p H\over\p J_z}
+N_\phi{\p H\over\p J_\phi}.
\]
 so {\it TorusGenerator} has a method {\tt interpeTorus}  that returns 
any  {\tt eTorus} that lies in this manifold.  
Let {\tt eTs} be a {\tt std::vector<eTorus>} that populates the resonant
manifold, and {\tt xs} be a {\tt std::vector<double>} that gives the
corresponding values of some quantity, for example $J_r$, that varies along
the manifold. Then any eTorus in the manifold can be constructed by executing
for some value of {\tt x} the line
   {\obeylines\tt\parindent=10pt 
eTorus eT(TG.interpeTorus(x, xs, eTs));
}
\noindent
 This version of {\tt interpeTorus} finds $i$ such that
$xs[i]\le x\le xs[i+1]$ and then calls {\tt TG.interpeTorus(x,eTs[i],eTs[i+1])}.
The corresponding method {\tt interpTorus(x,xs,Ts)} exists for plain tori.

\begin{figure}
\centerline{\includegraphics[width=.9\hsize]{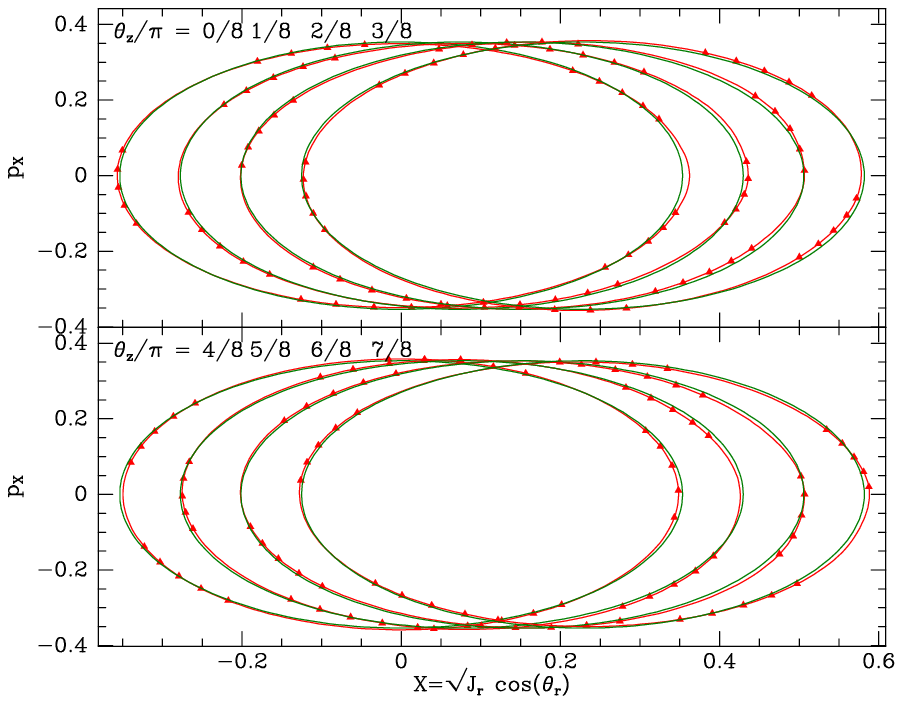}}
\centerline{\includegraphics[width=.9\hsize]{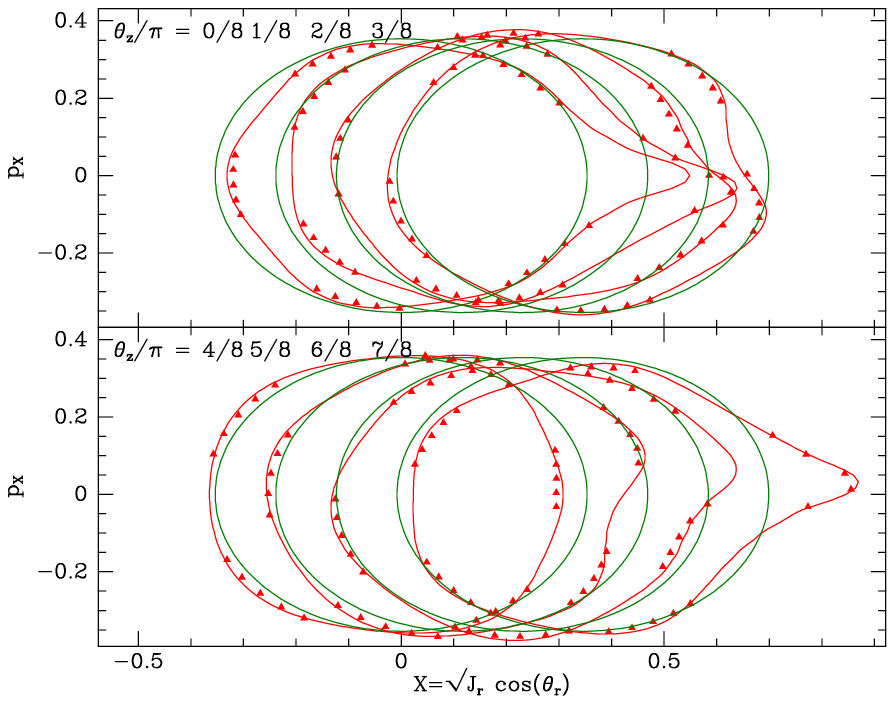}}
\caption{Cross sections $(\JT_r,\thetaT_r)$ through the torus of
Figs.~\ref{fig:t-seqs} to \ref{fig:Xsecs} at eight
different values of $\thetaT_z$ in the interval $[0,7/8]\pi$. In each panel
the data are shifted to the right by $\sim0.2$ of the maximum value of $X$ 
to minimise overlap between data for different values of $\thetaT_z$. The red
triangles are consequents from Runge-Kutta integration. They lie on the red curves drawn by the torus. The green 
ellipses show the corresponding cross sections produced by the underlying toy
map. Pericentre lies on the right and apocentre on the left. In the case of the lower pair
of panels, the torus was constructed with no Fourier coefficients in its toy
map, so all the work had to be done by the generating function
$S(\vthetaT,\vJ)$.}\label{fig:JrSoS}
\end{figure}

\subsection{Diagnostic plots}\label{sec:diagnostics}

Surfaces of section at constant values of a toy variable, for example
$\thetaT_z$, illustrate well the work done by the generating
function.  Fig.~\ref{fig:JrSoS} shows $(\JT_r,\thetaT_r)$ sections of
the torus that generated Figs.~\ref{fig:t-seqs} to \ref{fig:Xsecs} at
eight values of $\thetaT_z$ displayed  via the Cartesian coordinates
\[\label{eq:defX}
X=\sqrt{J^{\rm T}_r}\cos\thetaT_r \quad;\quad
p_X=\sqrt{J^{\rm T}_r}\sin\thetaT_r.
\]
 The red triangles are obtained by converting the coordinates $(R,z,\ldots)$
along the orbit into toy
angle-action variables and plotting a triangle when $\thetaT_z$ passes
$n\pi/8$ for $n=0,\ldots,7$. The triangles lie on the red curves that were
drawn by the torus, while the ellipses $\JT_r=\hbox{constant}$ are plotted in
green. Deviations between the red and green curves indicate the work done by
the generating function $S(\vJ,\thetaT)$. After execution of the lines
 {\obeylines\tt\parindent=10pt 
	std::vector<double> Xs, pXs;
	double thetaz = M\_PI/4, Xmax, pXmax;
	T.SoSthetaz(Xs, pXs, thetaz, 200, Xmax, pXmax);
}

\noindent
 {\tt Xs} and {\tt pXs} contain 200 points that delineate a red curve for
$\theta_z=\pi/4$ in Fig.~\ref{fig:JrSoS}. {\tt Xmax} and {\tt pXmax} contain
the largest absolute values of $X$ and $p_X$ around the curve. The
corresponding green ellipse can be obtained from the torus formed by the {\tt
T}'s toy map acting alone -- this is created by executing
 {\obeylines\tt\parindent=10pt 
	Torus BT(TG.fitBaseTorus(J));
} 
 \noindent the toy map of {\tt BT} has been optimised but all its $S_\vk$
vanish.

The upper pair of panels in Fig.~\ref{fig:JrSoS} was constructed from a
normal torus while the lower pair was constructed by a torus that had no
Fourier coefficients (eqns \ref{eq:def_xn} and \ref{eq:vFourier}) in its toy
map. As a consequence, in the lower panels the torus furnished by the toy map, plotted in green,
deviates considerably from the orbital torus plotted in red, while 444 $S_\vk$
were required to distort the green torus into the red one. By contrast, the torus shown in
the upper panels has only  114 $S_\vk$ and its \rms\
variation in $H$ is smaller by a factor of 33.

\section{An interface with observational data}\label{sec:obs}

Here we explain the contents of \agamaTwo's {\tt obs} namespace.

Models are created in galactocentric coordinates while observational data is
given in heliocentric coordinates. Positions on the sky, whether specified in
right-ascension and declination $(\alpha,\delta)$ or in Galactic longitude
and latitude $(\ell,b)$, are packaged into a struct {\tt PosSky}. This struct
includes a bool {\tt is\_ra} that's set true if the coordinates are
$(\alpha,\delta)$ and false otherwise. So we write
 {\obeylines\tt\parindent=10pt 
PosSky ad(30,40,true);
PosSky lb(30,40);
}
\noindent where the second declaration doesn't need a third argument because
{\tt is\_ad} is false by default. Angles are input and stored in degrees
rather than radians. The members of  {\tt PosSky} are {\tt l} and {\tt b}
regardless of the value of {\tt is\_ra}, so after the above line {\tt ad.b}
equals $40\,$deg.

Proper motions, in $\masyr$, are stored in a struct {\tt VelSky} that likewise
has a bool {\tt is\_ra}. Proper motion in the $\alpha$ direction is
understood to be $\mu_\alpha=\dot\alpha\cos\delta$ and similarly for
$\mu_\ell$. Positions and proper motions are stored in a struct {\tt
PosVelSky}, which has two components, {\tt pos} and {\tt pm}. Hence we might
write
 {\obeylines\tt\parindent=10pt 
VelSky muad(4.1,0.2,true);
PosVelSky pvs(ad,muad);
}
\noindent We could have achieved the same result by writing
 {\obeylines\tt\parindent=10pt 
PosVelSky pvs(30,40,4.1,0.2,true);
}

Transfer between equatorial and Galactic coordinates are done by functions
{\tt to\_RAdec} and {\tt from\_RAdec}. For example, after the call
 {\obeylines\tt\parindent=10pt 
PosVelSky lbv(from\_RAdec(pvs));
}
\noindent {\tt lbv.pos.b} will be the Galactic latitude of
$(\alpha,\delta)=(30,40)$ and {\tt lbv.pm.mub} will be the proper motion in
$b$. Attempts to mix positions in equatorial coordinates with proper motions
in Galactic coordinates, or to transfer coordinates into their own system
provoke error messages.

An instance of the class {\tt solarShifter} is used to move between
heliocentric coordinates and phase-space locations. An instance contains
the adopted phase-space location of the Sun. 
If we execute
 {\obeylines\tt\parindent=10pt 
solarShifter sun(intUnits);
}\noindent the Sun will be taken to be at
\begin{align}
(x,y,z)&=(-8.27, 0, 0.025)\kpc\cr
(v_x,v_y,v_z)&=(14, 251.3, 7)\kms
\end{align}
A different location can be set by instead executing
 {\obeylines\tt\parindent=10pt 
solarShifter sun(intUnits,\&posVelSun);
}\noindent
where {\tt posVelSun} is something like
 {\obeylines\tt\parindent=10pt 
coord::PosVelCar posVelSun(-8*intUnits.from\_Kpc,
\qquad ..,..,12*intUnits.from\_kms,..,..);
}\noindent
Given a distance $s$ (in kpc) and a line-of-sight velocity $V_{\rm los}$ (in
$\kms$) we obtain a star's phase-space location from
 {\obeylines\tt\parindent=10pt 
coord::PosVelCar xv(sun.toCar(lb,s,mulb,Vlos));
}\noindent
Sky positions can be specified in either equatorial or Galactic coordinates
-- the calculations are carried out in Galactic coordinates preceded by a
call to {\tt from\_RAdec} if necessary. Positions in cylindrical coordinates
can be obtained by replacing {\tt toCar} with {\tt toCyl}. These functions will
also process positions alone as in
 {\obeylines\tt\parindent=10pt 
coord::PosCyl Rz(sun.toCyl(lb,s));
}\noindent
where {\tt s} is the distance in kpc. 

To obtain sky coordinates  from
phase-space locations we use {\tt toSky}: after executing
 {\obeylines\tt\parindent=10pt 
PosSky lb(sun.toSky(p,s))
}\noindent
where {\tt p} is a phase-space location (in either Cartesian or cylindrical
coordinates), {\tt lb} will contain the star's Galactic coordinates (in
degrees) and {\tt s} will be its distance (in kpc). If we just want the
distance, we can execute {\tt sun.sKpc(p);}. To obtain complete heliocentric
kinematics, we execute
  {\obeylines\tt\parindent=10pt 
PosVelSky pvs(sun.toSky(p,s,Vlos));
}\noindent
The component of the Sun's velocity towards a sky location is obtained by
executing
  {\obeylines\tt\parindent=10pt 
double V=sun.Vreflex(lb);
}\noindent
If we know the star's distance, we can obtain the proper motion and
line-of-sight velocity  that would be
measured if the Sun were stationary in the Galaxy's rest frame from
  {\obeylines\tt\parindent=10pt 
VelSky vs(sun.Vreflex(lb, s, Vlos));
}

\subsection{Lines of sight}

Spectroscopic surveys probe both our Galaxy and external galaxies down a
relatively small number of lines of sight (\los), so it's important to be able to
extract from a galaxy model predictions for the distribution of stellar
observables along a specified \los. To this end \agamaTwo\ has classes
{\tt sunLos} and {\tt extLos}, both of which inherit from the class {\tt
BaseLos}. For these classes internal units are used to facilitate their use
in the sampling routines of the {\it galaxymodel} namespace.

To create a {\tt sunLos}, we execute
  {\obeylines\tt\parindent=10pt 
sunLos mylos(lb,sun);
}\noindent
where {\tt lb} is a {\tt PosSky} and {\tt sun} is  a {\tt solarShifter}. To
get the los through some location, replace {\tt lb} by the location's {\tt
coord::PosCar}. To create an {\tt extLos} we execute
  {\obeylines\tt\parindent=10pt 
extLos mylos(intUnits,x,y,incl,D);
}\noindent
 Here {\tt x,y} specify the Cartesian coordinates on the sky (in internal
units) of the
\los\ relative to the galaxy's projected centre, {\tt incl} (deg) is the galaxy's inclination,
and {\tt D} is its distance in internal units. The $x$ axis is assumed to coincide with the galaxy's line of
nodes (apparent major axis).

An optional final parameter to these creators is a pointer to a {\tt dustModel}
(described below). If a dust model is specified, the creator tabulates the
column density $N(s)$ as a function of distance $s$ so that extinctions $A_i$
in the $V,B,R,H$ and $K$ wavebands  can be obtained by calling the methods
{\tt A\_V(s)} etc. For example
  {\obeylines\tt\parindent=10pt 
double Av=mylos.A\_V(s);
}\noindent
yields the visual-band extinction to a point $s$ down the \los.

The method {\tt s(p)}, where {\tt p} gives the Cartesian coordinates of a
point, returns the distance (in internal units) to the Sun in the case of a {\tt sunLos} while in
the case of an {\tt extLos} it returns the (signed) distance
from where the
\los\ punches through the galaxy's $xz$ plane -- this point is well defined
except when {\tt incl=0} and the galaxy is face-on. The class {\it extLos} has a
method {\tt sVlos(p)} that returns both the distance and
line-of-sight velocity
of the {\tt PosVelCyl p}. For example, after execution of
  {\obeylines\tt\parindent=10pt 
std::pair<double,double> sV(mylos.sVlos(p));
}\noindent
{\tt sV.first} will contain the distance to {\tt p} while {\tt sV.second}
will contain {\tt p}'s velocity along the \los, both in internal units.

Both types of \los\ have a method {\tt deepest()}, which returns the point
where the \los\ comes closest to the galactic centre and a method {\tt
xyz(s)} that returns the Cartesian coordinates of the point distance $s$
(internal units) down
the \los.

The {\it galaxymodel} namespace provides several functions that sample a
\los. For example, {\it sampleLOS} returns as a vector of {\tt PosVelCyl} the
phase-space coordinates of stars along that line of sight, optionally sampled
within some magnitude interval. {\it sampleLOSsVlos} samples a \los\
similarly but returns only distances and $V_{\rm los}$ values. {\it
computeMomentsLOS} returns the projected density and the first and second
moments of the velocity distribution along a \los. With no \los\ specified,
{\it sampleVelocity} returns the velocities of a sample of stars at a
location. By specifying a \los, this sample can be restricted to stars that
have apparent magnitudes in a specified range.

\subsection{Dust}

A minimal dust model can be created by executing 
  {\obeylines\tt\parindent=10pt 
BaseDustModel(Rd,zd,rho0,Rw,Hw,from\_Kpc);
}\noindent
which creates a warped double-exponential dust layer. {\tt Rd}, {\tt zd} and
{\tt rho0} specify the scale lengths (in kpc) and the density at the origin.
{\tt Rw} nd {\tt Hw} specify a warp: the dust layer is centred on the
equatorial plane at $R<R_{\rm w}$ but at larger $R$ it's centred on
\[
z_{\rm w}=\Big({R\over R_{\rm w}}-1\Big)H_{\rm w}\sin\phi.
\]
A dust model can also be created by executing
  {\obeylines\tt\parindent=10pt 
BaseDustModel(ptr, rho0, from\_Kpc);
}\noindent
where {\tt ptr} is a pointer to a user-specified function: the density at
{\tt p} is {\tt rho0 * ptr->density(p)}.

More complicated dust models can be created by adding any number of spirals,
\cite{JT1966} cloud wakes 
and blobs to a {\tt BaseDustModel}.

\end{document}